\documentclass{aa7.0}

\usepackage{graphicx}
\usepackage[varg]{txfonts}
\usepackage{bm}

\usepackage{natbib}
\bibpunct{(}{)}{;}{a}{}{,}

\usepackage{hyperref}

\newcommand{\beq}{\begin{equation}}
\newcommand{\eeq}{\end{equation}}
\newcommand{\bea}{\begin{eqnarray}}
\newcommand{\eea}{\end{eqnarray}}
\newcommand{\req}[1]{Eq.~(\ref{#1})}

\newcommand{\am}{a_\mathrm{m}}
\newcommand{\dd}{\mathrm{d}}
\newcommand{\ee}{\mathrm{e}}
\newcommand{\EF}[1]{E_{\mathrm{F}#1}}
\newcommand{\gcc}{\mbox{g cm$^{-3}$}}
\newcommand{\kB}{k_\mathrm{B}}
\newcommand{\mel}{m_\ee}
\newcommand{\nbar}{\bar{n}}
\newcommand{\omg}{\omega_\mathrm{g}}
\newcommand{\nn}{\mathrm{n}}
\newcommand{\pF}[1]{p_{\mathrm{F}#1}}
\newcommand{\pp}{\mathrm{p}}
\newcommand{\Tcrit}{T_\mathrm{crit}}
\newcommand{\Teff}{T_\mathrm{eff}}

\begin{document}

\title{Magnetic neutron star cooling and microphysics}
                                                         
\author{
A. Y. Potekhin\inst{1,2,3}\thanks{\email{palex@astro.ioffe.ru}}
\and
G. Chabrier\inst{1,4}
}
\institute{
Centre de Recherche Astrophysique de Lyon, Universit\'e de
Lyon, Universit\'e Lyon 1, Observatoire de Lyon, Ecole Normale
Sup\'erieure de Lyon, CNRS, UMR 5574, 46 all\'ee d'Italie,
69364, Lyon Cedex 07, France
\and
Ioffe Institute,
Politekhnicheskaya 26, 194021, Saint Petersburg, Russia
\and
Central Astronomical Observatory at Pulkovo,
Pulkovskoe Shosse 65, 196140, Saint Petersburg, Russia
\and
School of Physics, University of Exeter, Exeter, UK EX4 4QL
}

\date{Received 31 August 2017 / Accepted 24 October 2017}

\abstract
{}{We study the relative importance of several recent updates of
microphysics input to the neutron star cooling theory and the effects
brought about by superstrong magnetic fields of magnetars, including the
effects of the Landau quantization in their crusts. 
}{We use a finite-difference code for simulation of neutron-star thermal
evolution on timescales from hours to megayears
with an updated microphysics input. 
The consideration of short timescales ($\lesssim1$ yr) is made
possible by a treatment of the heat-blanketing envelope without the
quasistationary approximation inherent to its treatment in traditional
neutron-star cooling codes.
For the
strongly magnetized neutron stars, we take into account the effects of
Landau quantization on thermodynamic functions and thermal
conductivities. We simulate cooling of ordinary neutron stars and
magnetars with non-accreted and accreted crusts and compare the results
with observations.
}{Suppression of radiative and conductive opacities in strongly
quantizing magnetic fields and formation of a condensed radiating
surface substantially enhance the photon luminosity at early ages,
making the life of magnetars brighter but shorter. These effects
together with the effect of strong proton superfluidity, which slows
down the cooling of kiloyear-aged neutron stars, can explain  thermal
luminosities of about a half of magnetars without invoking heating
mechanisms. Observed thermal luminosities of other magnetars are still
higher than theoretical predictions, which implies heating, but the
effects of quantizing magnetic fields and
baryon superfluidity help to reduce
the discrepancy.
}{}


\maketitle

\section{Introduction}
\label{sect:intro}

Heat stored in neutron stars after their birth is gradually lost to
neutrino emission and surface radiation. The rate of these losses
depends on the characteristics of a neutron star: its mass, magnetic
field, and composition of heat-blanketing envelopes. Theoretically
calculated luminosity of the star as a function of its age (the cooling
curve) is sensitive to the details of the theory of matter in extreme
conditions: high densities, temperatures, and magnetic fields, not
reachable in the terrestrial laboratory. Therefore, a comparison of
observed surface luminosity with theoretical predictions can potentially
provide information not only on the characteristics of a given star, but
also on the poorly known properties of matter in extreme conditions. The
theory of neutron star cooling has been developed in many papers,
starting from the pioneering works by \citet{ChiuSalpeter64} and
\citet{TsurutaCameron66}. For a recent review of this theory, see
\citet{PotekhinPP15}.

Thermal or thermal-like radiation has been detected from several classes
of neutron stars. Particularly interesting are isolated neutron stars
with confirmed thermal emission, whose thermal X-ray spectra are not
blended with emission from accreting matter or magnetosphere. A
comprehensive list of such objects found before 2014 has been compiled
by \citet{Vigano_ea13}. Many neutron stars from this list are
satisfactorily explained by the cooling theory (see, e.g.,
\citealp{Yakovlev_ea08,Vigano_ea13}, and Sect.~\ref{sect:nonmag}
below). However, the magnetars, a special class of neutron star with
superstrong magnetic fields (see
\citealp{MereghettiPM15,TurollaZW15,KaspiBeloborodov17}, for recent
reviews), are much more luminous than ordinary cooling neutron stars of
the same age. For example, at age $t\sim10$ kyr, typical thermal
luminosities are $\sim(10^{31}-10^{33}$) erg s$^{-1}$ for ordinary neutron
stars and several times ($10^{33}-10^{35}$) erg s$^{-1}$ for magnetars. Their high
luminosities are believed to be fed by magnetic energy stored in the
neutron star \citep{DuncanThompson92}, but the mechanism of the release
of this energy is uncertain. Several possible heating mechanisms
suggested by different authors have recently been analyzed by
\citet{BeloborodovLi16}.  

The characteristic magnetic fields of magnetars, determined from their
spindown according to the standard model of a rotating magnetic dipole
in vacuum, range from $\sim10^{13}$~G to $\approx2\times10^{15}$~G 
\citep{OlausenKaspi14}. In addition to the poloidal magnetic field at
the surface, the magnetars are believed to have much stronger toroidal
magnetic field embedded in deeper layers
\citep{GeppertKP06,PerezAzorinMP06}, which is
corroborated by observations of free precession of the magnetars
\citep{Makishima_ea14,Makishima_ea16}. For a characteristic poloidal
component $B_\mathrm{pol}$ of a neutron-star magnetic field to be
stable, a toroidal component $B_\mathrm{tor}$ must be present, such
that, by order of magnitude, $B_\mathrm{pol} \lesssim B_\mathrm{tor}
\lesssim 10^{16}\mbox{~G}\,({B_\mathrm{pol}/10^{13}\mbox{~G}})^{1/2}$
\citep{Akgun_ea13}. Moreover, theoretical arguments
\citep[e.g.,][]{BonannoUB05}, numerical simulations
\citep[e.g.,][]{Braithwaite08,Lasky_ea11,KiuchiYS11,Mosta_ea15}, and observational evidence
\citep[e.g.,][]{Tiengo_ea14} show that magnetars possess highly tangled,
small-scale magnetic fields, which can be several times stronger than
their dipole component (see also
\citealp{MereghettiPM15,LinkvanEysden16}, and references therein).

In this paper, we analyze the relative importance of different microphysics
inputs in cooling of ordinary neutron stars and magnetars.  We
model both neutron stars with ground-state and accreted envelopes. We
study the sensitivity of the cooling to different microphysics
ingredients of the theory, including the equation of state (EoS), baryon
superfluidity, opacities, and neutrino emission mechanisms, putting
emphasis on the role of recent updates of corresponding theoretical
models. We demonstrate the importance of the effects of Landau
quantization in the crust of the magnetars on the
EoS and thermal conductivity tensor, as well as formation of a
condensed radiating surface, possibly covered by a dense atmosphere. The
latter effects make the heat-blanketing envelope much more transparent
than predicted by the classical theory.

The paper is organized as follows. In Sect.~\ref{sect:equations}, we
state the main assumptions, recall the basic equations of the cooling
theory, and describe our numerical cooling code. In
Sect.~\ref{sect:physics} we briefly recall the essential physics of
neutron star cooling and summarize physics input. In
Sect.~\ref{sect:nonmag}, we examine the influence of different physics
input ingredients on cooling of nonmagnetized neutron stars, with special attention being paid to several modern microphysics updates.
Section~\ref{sect:magnetars} is devoted to the cooling of strongly
magnetized neutron stars and the role of Landau quantization. 
Conclusions are summarized in Sect.~\ref{sect:concl}. Appendices
\ref{sect:eosfits} and \ref{sect:F_B} specify some details of our
treatment of the EoSs of nonmagnetic and strongly magnetized neutron
stars.

\section{Simulation of thermal evolution}
\label{sect:equations}

The most massive central part of a neutron star, its core,
consists of a uniform mixture of baryons and leptons. The core is
surrounded by the crust, where atomic nuclei form a crystalline lattice,
and the electrons form almost ideal, strongly degenerate Fermi gas. In
an inner part of the crust, the lattice of nuclei is immersed in a
liquid of ``dripped'' (quasi-free) neutrons. The crust is covered by a
liquid layer, the ocean, where the lattice is destroyed by thermal
fluctuations. As the star cools down, the bottom layers of the ocean
freeze into the crust. The ocean, in turn, is usually covered by a
gaseous atmosphere, where the spectrum of outgoing thermal radiation is
formed.

About a day after the birth of a neutron star, the
temperature everywhere in its interior drops below
$10^{10}$~K. At this temperature, the chemical potentials
of nucleons at densities $\rho\gtrsim10^{12}$ \gcc{} and of
electrons at $\rho\gtrsim10^8$ \gcc{} (without the rest
energies) are much higher than their kinetic thermal
energies. At these conditions, a good approximation is to
describe the state of matter as cold nuclear matter in beta equilibrium,
resulting in an effectively barotropic EoS. Therefore neutron-star
cooling simulations are traditionally performed by treating the
internal mechanical structure of the star as
decoupled from its thermal structure. The outer boundary condition for
simulations of thermal evolution is then provided by a solution of the
stationary heat transport equation at $\rho<\rho_\mathrm{b}$, where the
commonly accepted value $\rho_\mathrm{b}=10^{10}$ \gcc{}
\citep{GudmundssonPE83} is a trade-off between the applicability of a
barotropic EoS at $\rho>\rho_\mathrm{b}$ and the stationary
approximation at $\rho<\rho_\mathrm{b}$.

This approach simplifies the cooling simulations, but it
does not allow one to trace the rapid processes in the
crust, accompanied by sudden heat release, which may result
from accretion episodes, starquakes, magnetic reconnection
events, and so on, and which are probably common in the magnetars
\citep{MereghettiPM15,TurollaZW15,KaspiBeloborodov17}. In
this paper we follow a different approach, which is
traditional for the nondegenerate, nonrelativistic stars,
and which was extended to the case of general relativity by
\citet{RichardsonVHS79}.

We assume the mechanical structure to be spherical. Appreciable
deviations from the spherical symmetry can be caused by ultra-strong
magnetic fields ($B\gtrsim10^{17}$~G) or by rotation with ultra-short
periods (less than a few milliseconds), but we will not consider such
extreme cases. We also assume a spherically symmetric thermal structure.
This assumption can be violated by local heat releases or by large-scale
(e.g., dipolar) strong magnetic fields. For highly tangled small-scale
fields, the spherical symmetry is a reasonable first approximation.

In the spherical symmetry, the thermal and mechanical structure are
governed by six first-order differential equations for radius $r$,
gravitational potential $\Phi$, gravitational mass $M_r$ inside a sphere
of radius $r$, luminosity $L_r$ passing through this sphere, pressure $P$,
and temperature $T$ as functions of the baryon number $a$ interior to a
given shell (\citealp{RichardsonVHS79}; cf.~\citealp{Thorne77}). The
time $t$ enters only the equation for $L_r$ in the form of the coordinate
time derivative of the entropy per baryon. 

The mechanical structure of the star is defined by equations
\bea
  \frac{\dd r}{\dd a} &=& \frac{1}{4\pi r^2 \nbar}
     \,\left(1-\frac{2GM_r}{r c^2}\right)^{1/2},
\label{drda}
\\
   \frac{\dd M_r}{\dd a} &=& \frac{\rho}\nbar\,
     \left(1-\frac{2GM_r}{r c^2}\right)^{1/2},
\\
   \frac{\dd\Phi}{\dd a} &=& G\,\frac{M_r+4\pi r^3 P/c^2}{
         4\pi r^4\nbar}\,\left(1-\frac{2GM_r}{r c^2}\right)^{-1/2},
\label{Phi}
\\
   \frac{\dd P}{\dd a}&=& -\left(\rho+\frac{P}{c^2}\right)
         \,\frac{\dd\Phi}{\dd a},
\label{dPda}
\eea
where $\nbar$ is the mean number density of baryons, $G$ is the Newtonian
constant of gravitation and $c$ is the speed of light in vacuum. These
equations are integrated from $r=0$ and $M_r=0$ at the center of the star
outwards, starting from a predefined central baryon density and keeping
$\rho$ and $P$ related to $\nbar$ via the EoS, until the outer boundary
condition is satisfied (e.g., a predefined mass density at the
outer boundary $\rho_\mathrm{b}$ is reached). The boundary
condition for $\Phi$ is provided by the Schwarzschild metric outside the
star,
\beq
   \ee^{2{\Phi(R)/c^2}} = 1-2GM/c^2R,
\label{PhiR}
\eeq
where $R$ and $M=M_R$ are the stellar radius and mass. In
practice, \req{Phi} is integrated for a shifted potential
$\Phi(r)-\Phi(0)$, with the initial value equal to zero at
the center of the star, and afterwards the value of the
shift $\Phi(0)$ is found from \req{PhiR}.

The heat flux through a spherical surface is related to gradient of the
redshifted temperature $\tilde{T}=\ee^{\Phi/c^2}T$ by equation
\beq
   L_r = -(4\pi r^2)^2\,\nbar\,\kappa\,\ee^{-{\Phi/c^2}}\,
        \frac{\dd \tilde{T}}{\dd a},
\label{L_T}
\eeq
where $\kappa$ is the thermal conductivity measured in the local. 
Finally, time-dependence is introduced by equation
\citep{RichardsonVHS79}
\beq
\frac{\dd (L_r\ee^{2\Phi/c^2})}{\dd a} = 
   \ee^{2\Phi/c^2}\left(\mathcal{E}
   -  T \,\ee^{-\Phi/c^2}
   \frac{\partial s}{\partial t} \right),
\label{Lda}
\eeq
where $\mathcal{E}$ is the net rate of energy generation per baryon
and $\partial s / \partial t$ is the
coordinate time derivative of the entropy per baryon ($\ee^{-\Phi/c^2}
{\partial s}/{\partial t}$ is the derivative evaluated in the local
rest frame of matter). Equation (\ref{Lda})
can
be combined with \req{L_T} to form
\beq
\frac{c_P}\nbar\,\ee^{\Phi/c^2}\,
\frac{\partial T}{\partial t} = 
     \frac{\partial}{\partial a}\,K(a)\,
         \frac{\partial\tilde{T}}{\partial a}
      + \frac{\tilde{H}-\tilde{Q}}{\nbar}.
\label{heat_diffusion1}
\eeq
Here, $c_P$ is the heat capacity per unit volume at constant pressure,
$\tilde{H}=\ee^{2{\Phi/c^2}}H$ and $\tilde{Q}=\ee^{2{\Phi/c^2}}Q$ are
redshifted powers of energy sources and sinks, respectively, per unit
volume, and
\beq
   K(a)\equiv (4\pi r^2)^2\, \nbar\, \kappa\, \ee^{\Phi/c^2}.
\eeq
Equation (\ref{heat_diffusion1}) can be written in the form of 
the usual thermal diffusion equation
\beq
\frac{c_P}\nbar\,
\frac{\partial\tilde{T}}{\partial t} = 
     \frac{\partial}{\partial a}\,K(a)\,
         \frac{\partial\tilde{T}}{\partial a}
      + \frac{\tilde{H}-\tilde{Q}}\nbar
      + \frac{c_P}\nbar\,\tilde{T}\,\frac{\partial\Phi}{c^2\partial t}.
\label{heat_diffusion}
\eeq
In practice, the last term on the right-hand side is small 
compared to typical values of the left-hand side. 
Therefore, in a finite-difference scheme of solution, this term can be
treated as an external source, with $\partial\Phi/\partial t$ evaluated
from the solution at the preceding time step.
The boundary condition to \req{heat_diffusion} at the stellar center is
$\partial\tilde{T}/\partial a =0$. The outer boundary
condition follows from \req{L_T}: 
\beq
\left.
\frac{\partial\tilde{T}}{\partial a}
\right|_{a=a_\mathrm{b}}
 = -\frac{\ee^{{\Phi/c^2}}\,{L}_\mathrm{b}}{(4\pi
r^2)^2\nbar\kappa}, 
\label{L_b}
\eeq 
where ${L}_\mathrm{b}$ is the energy flux through the outer boundary
$a=a_\mathrm{b}$,
which is provided by the quasi-stationary thermal structure of a thin
envelope outside this boundary. Since we solve the nonstationary problem using the
temperature-dependent EoS in the outer crust, the position of the
boundary is not restricted by the requirement that the
plasma should be degenerate at
$\rho_\mathrm{b}=\rho(a_\mathrm{b})$.
Therefore, the thickness of the quasi-stationary envelope can be
adapted to an astrophysical problem of interest. In the present work, we
choose  the mass of the quasi-stationary envelope $\Delta M$ so as to ensure that
plasma is fully ionized at $\rho>\rho_\mathrm{b}$. At $B=0$, this
condition is guaranteed for the mass of an  envelope $\Delta
M=10^{-12}M_\odot$. In the case of strong magnetic fields we find it
appropriate to increase the envelope mass to $\Delta
M=(10^{-11}-10^{-9})M_\odot$, because the Landau quantization
suppresses opacities (see Sect.~\ref{sect:opacities}), which results in
quick thermal relaxation of such massive envelopes. We take the partial
ionization into account in the outer quasi-stationary boundary layer
only. In the deeper layers, where the time-dependent problem is being
solved, we use the fully ionized plasma model.

We solve the set of equations (\ref{drda})\,--\,(\ref{heat_diffusion})
by a finite-difference scheme on a non-uniform, adaptive grid in $a$ and
$t$. First, at $t=0$,  we define a starting temperature profile (usually
constant $\tilde{T}=10^{10}$~K, but we have checked that the
start from $\tilde{T}=5\times10^{10}$~K changes the surface
temperature by $<1$\% at $t\gtrsim1$~hr) and solve the set of equations
(\ref{drda})\,--\,(\ref{dPda}) by the Runge-Kutta method. 
In order to prevent accuracy loss in the outer crust and ocean, where
$a$ is nearly constant as a function of $\rho$, we use the difference
$(a_\mathrm{b}-a)$ as an independent variable.  At each $t=t_j$, we
introduce a nonuniform grid in this variable and choose a time step
$\Delta t_j$ so as to ensure smallness of variations of $T$ and $P$
between the neighboring grid points and between the time $t_j$ and
$t_{j+1}=t_j+\Delta t_j$. We solve the difference equation, that
approximates \req{heat_diffusion}, by a purely time-implicit
energy-conservative scheme (\citealp{Samarskii}, Chapter~8, Eqs.~(35),
(36), (39)). This scheme would be unconditionally stable if $c_P/\nbar$,
$K(a)$, $(\tilde{H}-\tilde{Q})/\nbar$ were constant, so the time step
$\Delta t_j$ is not limited by the Courant-Friedrichs-Lewy condition.
The values of the coefficients are evaluated at the next layer
$t_{j+1}$, so the scheme is nonlinear. Self-consistent solution for 
$\tilde{T}^{j+1}\equiv \tilde{T}(t_{j+1})$ and coefficients of the 
difference equation at $t=t_{j+1}$ is found
by an iterative method. At each iteration, the coefficients of the
equation, found on the previous iteration, are kept fixed, so that the
scheme becomes linear and the values of $\tilde{T}^{j+1}$ on the new
layer $t_{j+1}$ are found by the elimination method in terms of the
function $\tilde{T}^j$ on the current layer $t_j$ \citep{Samarskii}. If
a variation of $\tilde{T}$ at the time step $\Delta t_j$ proves to be
insufficiently small at some point, we diminish $\Delta t_j$ and repeat
the calculation for a new layer $t_{j+1}$. After the solution is found
for $\tilde{T}$ on a given layer, the mechanical structure is adjusted
(if necessary) using Eqs.~(\ref{drda})\,--\,(\ref{dPda}). After the
adjustment, the values of $\tilde{T}$ on the new grid in $a$ are
obtained by interpolation from the values on the previous grid. The
overall accuracy of the solution has been checked by variation of the
criteria for choosing steps in $t$ and $a$ and the number of iterations.

\section{Physics input}
\label{sect:physics}

\subsection{The essential physics of neutron star cooling}
\label{sect:essential_physics}

The essential physics ingredients needed to build a model of a cooling
neutron star, are the EoS (including $P$, $\rho$, and $c_P$
as functions of $\nbar$ and $T$), rates of different neutrino emission
mechanisms responsible for the energy sink, and thermal conductivity.
These ingredients are substantially different in different shells of the
star: the atmosphere and outer ocean, where ionization of
atoms can be incomplete; the inner
ocean and outer crust, which consist of fully ionized electron-ion
Coulomb plasmas, liquid or solid depending on $\rho$, $T$, and chemical
composition; the inner crust at
$\rho_\mathrm{nd}<\rho<\rho_\mathrm{cc}$, where free neutrons are
present; and the core at $\rho>\rho_\mathrm{cc}$, consisting of a
uniform matter. Here,
$\rho_\mathrm{nd}\approx4.3\times10^{11}~\gcc$ is the neutron drip
density, and $\rho_\mathrm{cc}\approx(1.0-1.5)\times10^{14}$ \gcc{} is
the density at the crust-core phase transition.
For the present work, we have
restricted ourselves by the $npe\mu$ matter, that is, matter composed of
nucleons and leptons without free hyperons, mesons, or quarks.

The heat capacity, neutrino emissivity, and
heat transport by nucleons can be strongly affected by nucleon
superfluidity in the inner crust and the core, by
many-body \emph{in-medium} effects in the core, and by magnetic field in
any region of the star, if the field is sufficiently strong (see
\citealp{PotekhinPP15} for review).

In magnetic fields, the heat transport becomes anisotropic, so that the
conductivity is a tensor. For the dominant electron heat
conduction mechanisms, this effect is important if the Hall
magnetization parameter $\omg/\nu_\mathrm{coll}$ is large enough. Here,
$\omg$ is electron gyrofrequency and $\nu_\mathrm{coll}$ is an effective
collision frequency of heat carriers (for quantitative estimates in
different heat-transport regimes see, e.g., \citealp{PotekhinPP15}).
The scalar $\kappa$ that is needed in the spherically symmetric model,
Eqs.~(\ref{L_T})\,--\,(\ref{L_b}), is an appropriate
effective value. If the field is large-scale, for example dipolar, then
a locally one-dimensional approximation can be applied with effective
$\kappa=\kappa_\|\,\cos^2\theta_B+\kappa_\perp\,\sin^2\theta_B$, where
$\kappa_\|$ and $\kappa_\perp$ are the conductivities along and across
the field, respectively, and $\theta_B$ is the angle between the
magnetic field and the normal to the surface. In the case of highly
tangled magnetic fields, which we consider here, a more appropriate
model is the local average, $\kappa=\kappa_\|/3+2\kappa_\perp/3$
\citep[e.g.,][]{PotekhinUC05}.

In the $npe\mu$ matter of the core of a neutron star, heat
is carried mainly by electrons, muons, neutrons, and
protons. In the crust, the main heat carriers are electrons
(contributions from neutrons in the inner crust and from
phonons are less important). In the ocean and atmosphere,
the competing heat transport mechanisms are the radiative
and electron conduction.

If the magnetic field is nonquantizing, it does not affect heat
transport by uncharged carriers (photons, phonons, and neutrons). It
also does not affect the longitudinal transport by charged particles
(electrons, muons, and protons), but suppresses the corresponding
transverse heat transport. In the degenerate matter, the suppression
factor is $\approx1/[1+(\omg/\nu_\mathrm{coll})^2]$. The nonquantizing
magnetic field does not affect the EoS.

A quantizing magnetic field affects both longitudinal and transverse
conductivities and the EoS. If the field is strongly quantizing, so that
all particles reside on the ground Landau level, these effects are quite
pronounced. For the electrons in the crust, the field is strongly
quantizing if $T\ll T_\mathrm{cycl}=\hbar eB/\mel c \kB =
1.3434\times10^8\,B_{12}$~K and
$\rho<\rho_B$, where
\beq
   \rho_B=\frac{m_\mathrm{u}}{\pi^2\sqrt{2}\,Y_\ee}\,
   \left(\frac{eB}{\hbar c}\right)^{3/2} =
      7045\,Y_\ee^{-1}B_{12}^{3/2}~ \gcc,
\label{rho_B}
\eeq
$m_\mathrm{u}$ is the unified
atomic mass unit, $Y_\ee$ is the number of electrons per baryon, and
$B_{12}\equiv B/10^{12}$~G (see, e.g., \citealp{PC13}, Sect.~4.2.2). For
example, if the crust has the ground-state composition and
$B=10^{16}$~G, then $\rho_B\approx1.8\times10^{10}$ \gcc.
In practice, magnetic field can be considered as nonquantizing if
either $\rho\gg\rho_B$ or $T \gg T_\mathrm{cycl}$.

\subsection{Equation of state}
\label{sect:EoS}

The EoS for the core of neutron stars is sensitive to the details of
fundamental physical theory of matter at extreme densities. There is
thus an intrinsic connection between the macroscopic structure and
evolution of the neutron stars and the underlying fundamental
interactions between the constituent particles at the microscopic level.
There is a large number of different theoretical approaches to
construction of the EoS of superdense matter (see, e.g., the excellent
recent review by \citealp{Oertel_ea17}). 
Here we mainly use the results by
\citet{Pearson_ea17}, who have developed a unified treatment
of the outer and inner crusts and the core of a neutron star,
calculating the zero-temperature equation of state  in each region with
the same energy-density functional from the ``Brussels-Skyrme'' (BSk)
family of functionals. They considered  three such functionals,  labeled
BSk22, BSk24 and BSk25, which are based on generalized  Skyrme-type
forces supplemented with realistic contact pairing forces. The
parameters of these models are constrained by \citet{GorielyCP13} to fit
the database of 2353 nuclear masses \citep{AME2012} and to be consistent
with the EoS of neutron-star core calculated by \citet{LiSchulze08}
within the Brueckner-Hartree-Fock approach, using the realistic Argonne
v18 nucleon-nucleon potential \citep{WiringaSC95} and the
phenomenological three-body forces that employ the same meson-exchange
parameters as the Argonne v18 potential. We have selected to use the EoS
BSk24 as our basic model, because it provides the best agreement with
various experimental constraints  (nuclear mass measurements,
restrictions derived from heavy-ion collision experiments, etc.). It is
very similar to the BSk21 EoS model (\citealp{Potekhin_ea13} and
references therein), based on the generalized Skyrme functional fitted
to the previous atomic mass evaluation.

In view of a considerable theoretical uncertainty in EoS properties at
supranuclear densities $\rho\gtrsim\rho_0$, where 
$\rho_0\approx2.7\times10^{14}$ \gcc{} is the normal nuclear density, we also
use an alternative APR EoS \citep{AkmalPR98}, based on variational
calculations. We adopt the version of the APR model, named
A18+$\delta$v+UIX$^*$ in \citet{AkmalPR98}, where the Argonne v18
potential is supplemented by three-body force UIX$^*$ and so-called
relativistic boost interaction \citep{ForestPF95}. The force UIX$^*$ is
the phenomenological Urbana UIX three-body force model
\citep{Pudliner_ea95}, refitted to take account of the relativistic
boost. The APR EoS is not unified: it is applicable only to the core but
not to the crust. In the crust, therefore, we continue using the BSk24
model. For comparison, we also consider the ground-state nuclear
composition of the crust calculated in the
Hartree-Fock approximation by \citet{NegeleVautherin73} (hereafter NV)
and the EoS, calculated by \citet{DouchinHaensel01} using the liquid-drop model with parameters
derived from the Skyrme-Lyon effective potential SLy4.

Accretion of fresh material onto a neutron star can change the crust
composition. We include this possibility by using the model of
consecutive layers of H, $^4$He, $^{12}$C, and $^{16}$O, previously
employed in \citet{Potekhin_ea03}, but with more accurate $^{12}$C and
$^{16}$O boundaries \citep{PC12}, determined by the balance between the
cooling due to neutrino emission  and heating due to thermonuclear
burning. Beyond the $^{16}$O boundary, we adopt the composition of the
accreted crust obtained by \citet{HaenselZdunik90}.

Most of the quantities related to the inner crust and the core that are needed
for modeling thermal evolution of a nonmagnetized neutron star (pressure
and energy density, number fraction of electrons in the core,
characteristics of nuclei in the inner crust, effective masses of
nucleons, etc.) are implemented in our code via explicit
parametrizations (see Appendix~\ref{sect:eosfits}). They are based on
the theoretical models (BSk, APR, SLy4) which neglect the effects
of finite temperature and strong magnetic fields ($T=0$, $B=0$
approximation).

In the outer crust and the ocean, all thermodynamic functions at any $B$
and $T$ are provided by the model of a fully ionized magnetized Coulomb
plasma \citep{PC13}. This model is not directly applicable in the inner
crust and the core, because it does not take into account thermodynamic
effects of weak and strong interactions between free nucleons, which are
important at $\rho>\rho_\mathrm{nd}$. In this density range, we use the
nonmagnetic EoS models described above and add corrections due to the
magnetic field according to an approximate model described in
Appendix~\ref{sect:F_B}. The same model is used to calculate specific
heat contributions of all particles at all densities and magnetic fields
(with the use of effective masses of neutrons and protons, $m_\nn^*$ and
$m_\pp^*$).

Specific heat contributions of free baryons can be affected by their
superfluidity. The principal types of superfluidity arise from the
singlet Cooper pairing of protons in the core, triplet pairing of
neutrons in the core, and singlet pairing of free neutrons in the inner
crust of the star. The modifications of the heat capacity by these types
of superfluidity are described by reduction factors, presented
by \citet{YakovlevLS99} (with a typo fixed according to footnote~1 in
\citealp{PotekhinPP15}) as functions of $T/T_{\mathrm{crit}}$, where
$T_{\mathrm{crit}}$ is the critical temperature for a given type of
superfluidity. In the dense matter, different types of triplet pairs of
neutrons can form a superposition \citep{Leinson10}. For simplicity
hereafter we restrict ourselves by considering the $^3$P$_2$, $m_J=0$
superfluidity in the case of triplet pairing (``Type B superfluidity''
of \citealp{YakovlevLS99}).

The critical temperatures $T_{\mathrm{crit}}$ are related to the gaps in
the energy spectra of the baryons, which are sensitive to the details of
underlying microscopic theory. A wealth of models have been developed in
the literature, resulting in different dependencies $T_{\mathrm{crit}}$
on the number densities of neutrons $n_\nn$ and protons $n_\pp$. In
general, these dependencies have an umbrella shape, with a maximum typically
$\sim10^9-10^{10}$~K for singlet type and $\lesssim10^9$~K for triplet
type of the Cooper pairs. We adopt the gap parametrization of
\citet{KaminkerHY01} and take the parameter values from Table~1 of
\citet{Ho_ea15}.

\subsection{Heat transport}
\label{sect:opacities}

In the core of a neutron star, the heat is carried by baryons ($n$, $p$)
and leptons ($e$, $\mu$). This heat transport is sensitive to the baryon
superfluidity. The conduction by baryons is also affected by the
in-medium effects. 

The most advanced studies of these heat transport mechanisms with
allowance for the effects of superfluidity have been performed by
\citet{ShterninYakovlev07} (in the case of transport by leptons) and by
\citet{ShterninBH13} (in the case of transport by baryons). The results
of \citet{ShterninYakovlev07} are given by analytical fits, which we have
incorporated in the cooling code. The results of \citet{ShterninBH13}
are taken into account in an approximate manner, according to
\citet{PotekhinPP15}, who find it sufficient to multiply the
conductivities obtained in the effective-mass approximation
\citep{BaikoHY01} by a factor of 0.6 to reproduce the thermal
conductivity of \citet{ShterninBH13} with an accuracy of several percent
in the entire density range of interest.

The most important heat carriers in the crust and ocean of
the neutron star are the electrons. In the atmosphere, the heat is
carried mainly by photons. In general, the two mechanisms
work in parallel, hence
$
   \kappa=\kappa_\mathrm{r}+\kappa_\ee,
$
where $\kappa_\mathrm{r}$ and $\kappa_\ee$  denote
the radiative (r) and electron (e) components of the thermal
conductivity $\kappa$. 

We calculate the electron thermal conductivities in the crust and ocean,
including the effects of strong magnetic fields, as described in
\citet{PotekhinPP15}. The radiative conductivity is calculated in the
model of fully ionized plasma, taking into account free-free transitions
and scattering,  \beq \kappa_\mathrm{r} =
\frac{16\sigma_\mathrm{SB}T^3}{3\rho K_\mathrm{R}}, \eeq where
$K_\mathrm{R}$ is the Rosseland mean radiative opacity, and
$\sigma_\mathrm{SB}$ is the Stefan-Boltzmann constant. For
nondegenerate, nonrelativistic electron-ion plasma without a quantizing
magnetic field, the photon-electron scattering opacity $K_\mathrm{sc}$
equals $n_\ee\sigma_\mathrm{T}/\rho$, where $n_\ee$ is the number
density of electrons and $\sigma_\mathrm{T}$ is the Thomson cross
section. Under the same conditions, the Rosseland opacities due to the
free-free transitions $K_\mathrm{ff}$ and due to the combined action of 
both free-free transitions and scattering were calculated and fitted by
analytical formulae  \citep{PotekhinYakovlev01}, based on a fit to the
frequency-dependent Gaunt factor obtained by \cite{Hummer88}.

Quantizing magnetic fields reduce the Rosseland mean radiative
opacities. The reduction factors for the nondegenerate, nonrelativistic
plasmas were calculated by  \citet{SilantievYakovlev80} and fitted by
\citet{PotekhinYakovlev01}.

Propagation of electromagnetic waves is quenched at photon frequencies
below the plasma frequency. The resulting increase of the Rosseland
opacity can be approximated by a density-dependent multiplication
factor, which equals 1 at low $\rho$ and high $T$ and increases at
high $\rho$ or low $T$ \citep{Potekhin_ea03}.

The scattering opacities are modified by the electron degeneracy at
high $\rho$ and by the Compton effect at $T\gtrsim10^8$~K. An accurate
analytical description of both these effects is given by
\citet{Poutanen17}.

The free-free opacities are suppressed by electron degeneracy. In the
absence of the Landau quantization, the free-free opacities at arbitrary
degeneracy have been fitted by \citet{Schatz_ea99}, based on numerical
calculations of \citet{Itoh_ea91}. The latter calculations, as well as
the above-mentioned results of \citet{Hummer88}, were based on the
tables of free-free absorption coefficient  as a function of the
electron velocity and photon frequency, calculated by
\citet{KarzasLatter61}.

The fit of \citet{Schatz_ea99} is inapplicable in the case of quantizing
magnetic fields. On the other hand, the results of
\citet{SilantievYakovlev80} and \citet{PotekhinYakovlev01} are only
applicable for nondegenerate nonrelativistic plasmas. 
In practice, for the $T$ and $B$ values typical for neutron star crusts,
the effects of electron degeneracy are most important
at $\rho\gtrsim\rho_B$, where the field is only weakly
quantizing. Therefore we neglect the effect of quantizing magnetic field
on the radiative opacities of degenerate electrons and apply the
magnetically modified opacities for nondegenerate electrons. A smooth
transition between the two regimes is provided by the weight factor
$\exp(-T/T_{\mathrm{F,e}})$, where $T_{\mathrm{F,e}}$ is the electron
Fermi temperature in a strongly quantizing magnetic field \citep{PC13}.

At high temperatures $T\gtrsim10^9$~K,  electron-positron pairs
become abundant and contribute to the radiative opacities. We calculate
the abundance of the $e^+e^-$ pairs from the standard relation
$\mu_++\mu_-=0$ \citep{LaLiSP1}, where $\mu_\pm$ are the chemical
potentials of $e^\pm$, including the rest energy, calculated with
allowance for the quantizing magnetic field.

\begin{figure}
\centering
\includegraphics[width=\columnwidth]{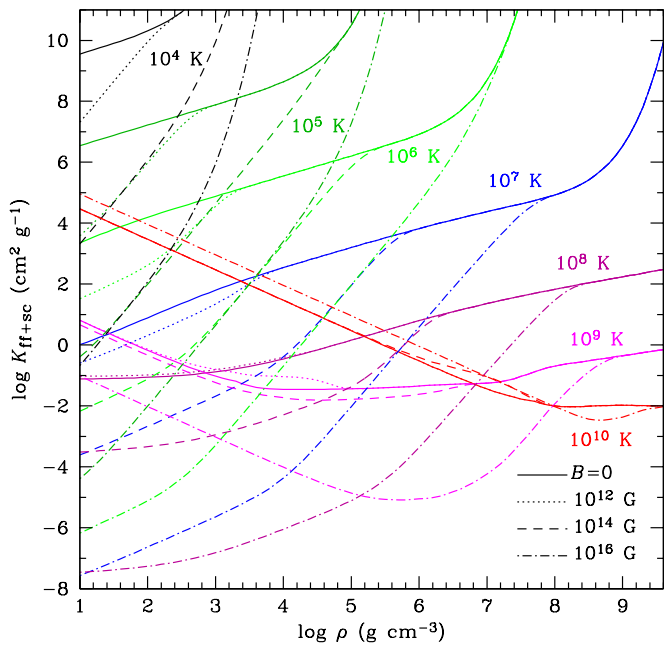}
\caption{Rosseland mean radiative opacities due to the free-free transitions and
Compton scattering in the model of fully ionized iron plasma as functions of mass
density at different temperatures (marked near the curves) and magnetic
fields $B=0$, $10^{12}$~G, $10^{14}$~G, and $10^{16}$~G (shown by
different line styles).
}
\label{fig:radopac}
\end{figure}

Figure~\ref{fig:radopac} shows examples of radiative opacities,
calculated at different temperatures, densities, and magnetic field
strengths in the model of a fully ionized iron plasma. We see that
strong magnetic fields reduce the opacities at lower densities by orders
of magnitude. At high densities, the plasma becomes degenerate, and the
opacities merge with nonmagnetic ones. The increase of the opacities at
small $\rho$ and high $T$ is due to the contribution of the $e^+e^-$
pairs. The sharp increase at high $\rho$ and at low $T$ is due to the
plasma-frequency cutoff. However, the electron conduction anyway
dominates at high $\rho$ and low $T$, so the cutoff is unimportant for
cooling simulations. If we replace iron by light chemical elements,
which is appropriate in the case of accreted envelopes, the opacities
decrease (e.g., for He the decrease reaches nearly an order of magnitude
at $T\lesssim10^8$~K), but the picture remains qualitatively the same.

In the atmosphere, which provides the outer boundary condition to
\req{heat_diffusion}, plasma can be partially ionized. For the
nonmagnetic atmospheres, we use the EoS and Rosseland mean opacities
provided either by the Opacity Library (\textsc{opal}, \citealp{OPAL})
or by the Opacity Project (\textsc{op}, \citealp{OP} and references
therein). We have checked that the differences between the \textsc{opal}
and \textsc{op} opacities are negligible for the conditions of our
interest. For strongly magnetized atmospheres, we use the model of fully
ionized plasma for iron and the EoS and opacities from \citet{PC04} for
hydrogen.

\subsection{Neutrino emission}
\label{sect:neutrino}

\citet{YKGH} reviewed the rich variety of reactions with neutrino
emission in the compact stars and presented convenient fitting formulae
for applications. The rates of reactions with participation of free
neutrons and protons are strongly changed if these particles are
superfluid. Moreover, the very existence of nucleon superfluidity gives
rise to a new neutrino emission mechanism by Cooper pair breaking and
formation (PBF), most powerful at $T\sim T_{\mathrm{crit}}$.

The most important reactions in the neutron-star crust and $npe$ matter
in the core with references to the appropriate fitting formulae are
collected in Table~1 of \citet{PotekhinPP15}.  In the crust, they are
plasmon decay, electron-nucleus bremsstrahlung, and electron-positron
annihilation. In the core, the most important reactions are different
branches of direct and modified Urca processes, baryon bremsstrahlung,
and the PBF at $T\sim \Tcrit$. The reactions with participation of
muons, fully analogous to those with the electrons, should be included
for the $npe\mu$ matter.  A strong magnetic field brings about
additional neutrino emission processes: electron synchrotron radiation
and, if protons are superconducting, bremsstrahlung due to scattering of
electrons on fluxoids \citep{YKGH}.

The direct Urca processes are most powerful, but operate only if the
proton fraction is large enough, which occurs above a certain threshold
baryon density $\nbar_\mathrm{DU}$ \citep[e.g.,][]{Haensel95}.
Therefore, those neutron stars whose mass $M$
exceeds a threshold $M_\mathrm{DU}$, where $\nbar$ exceeds
$\nbar_\mathrm{DU}$ at the center of the star, rapidly cool down via the
direct Urca processes \citep{Lattimer_ea91,PageApplegate92}. 

It is worthwhile noticing several updates in the relevant neutrino
reaction rates that have been developed recently, which improve the
results of \citet{YKGH}. An improved fit to the emission rate due to
plasmon decay, which has a larger validity range, has been constructed
by \citet{KantorGusakov07}. Electron-nucleus bremsstrahlung rates for
arbitrary (not only ground state) composition of the crust have been
calculated and fitted by \citet{OfengeimKY14}. The reduction of the
neutrino emissivity of modified Urca and nucleon-nucleon bremsstrahlung
processes by superfluidity of neutrons and protons in neutron-star cores
has been revised by \citet{Gusakov02}, who also presented a corrected
phase-space factor for the modified Urca process in the case of a large
proton fraction. Neutrino emission caused by the Cooper PBF has been
recalculated by accurately taking into account conservation of the
vector weak currents by \citet{LeinsonPerez06} for the singlet Cooper
pairing of baryons and by \citet{KolomeitsevVoskresensky08} for the
triplet pairing. \citet{Leinson10} described a modification of the PBF
neutrino emission rate due to the anomalous contribution to the axial
current, arising from the off-diagonal components of the vertex matrix
in the Nambu-Gorkov formalism for the nucleon interactions with the
external neutrino field.

Not all of these improvements have been so far accurately included in
the widely used neutron-star cooling codes and recent calculations
\citep[e.g.,][]{Page_ea09,Page_ea11,Ho_ea15,ShterninYakovlev15,Page16,Taranto_ea16}.
For example, the anomalous axial PBF contribution results in a
multiplication of the PBF neutrino-emission rate in the case of triplet
pairing by a factor of 0.19 compared to \citet{YKGH}, which is four
times smaller than the factor 0.76 effectively used in the recent
cooling calculations. Besides, the suppression of the contribution to
the PBF emission rate in the vector channel of weak interactions
\citep{LeinsonPerez06,KolomeitsevVoskresensky08} is usually taken into
account by setting this contribution to zero, whereas a more accurate
estimate is given by a small but non-zero multiplication factor $\sim
(\pF{}/m^*c)^2$, where $\pF{}$ is the Fermi momentum of the relevant
nucleons.

In the superdense matter of the core, the neutrino emission rates are
affected by the many-body ``in-medium effects'' (e.g.,
\citealp{Voskresensky01}, and references therein). We take account of
these effects following \citet{ShterninBaldo17}, who have obtained
in-medium enhancement factors for the modified Urca emission rates. They
also found an additional enhancement of the modified Urca emission rate
near the threshold for the opening of the direct Urca process, which
results in a non-standard temperature dependence of the emission rate
$Q\propto T^7$, intermediate between $Q\propto T^8$ and $T^6$ for the
modified and direct Urca processes, respectively.

Some neutrino emission rates can be modified by strong magnetic fields. 
For instance, \citet{BaikoYakovlev99} showed that the threshold for the
opening of the direct Urca processes is smeared out over some
$B$-dependent scale, and described this smearing by simple formulae,
which we include in the present treatment of the cooling. They also
showed that a strong magnetic field causes oscillations of the reaction
rate in the permitted domain of the direct Urca reaction, but the latter
effect, albeit interesting, appears to be unimportant for the cooling.

\begin{figure}
\centering
\includegraphics[width=\columnwidth]{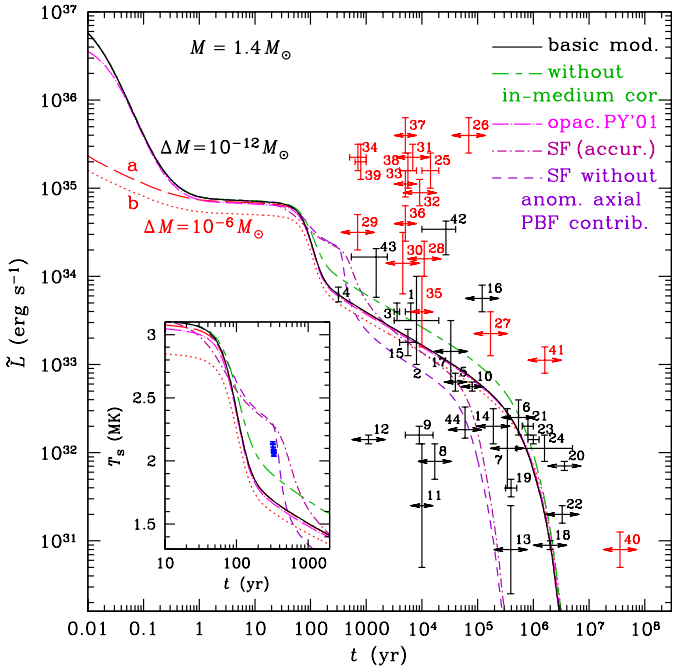}
\caption{Theoretical cooling curves (redshifted thermal luminosity
$\tilde{L}$ as a function of stellar age $t$) for a neutron star with mass
$M=1.4\,M_\odot$, calculated using different microphysics inputs (see
text for detail), versus observations of thermally emitting neutron
stars.  Vertical errorbars show estimated thermal luminosities,
horizontal errorbars are estimated ranges of kinematic ages, short
horizontal arrows replace the horizontal errorbars in the cases where no
confidence interval for the kinematic age is found in the literature,
and longer horizontal arrows are placed if no kinematic age is available
(in such cases, the characteristic ages are adopted). Numbers 1\,--\,41
enumerate the entries in Table~3 of \citet{PotekhinPP15}. We have
updated the data for object 4 according to \citet{Posselt_ea13} and
added objects 42 (XMMU J173203.3$-$344518,
\citealp{Klochkov_ea15}), 43 (CXOU J181852.0$-$150213,
\citealp{Klochkov_ea16}), and 44 (PSR J0633+0632,
\citealp{Danilenko_ea15,Karpova_ea17}). The errorbars and
arrows for magnetars are drawn in red color. The inset shows the
nonredshifted effective surface temperature $\Teff$ as function of $t$ in
a shorter time interval, which corresponds to the ages at which nucleon
superfluidity is expected to develop in the interior of the neutron
star. The symbols in the inset reproduce the data for the central
compact object in the Cas A supernova remnant from Table~I of
\citet{Ho_ea15}. The solid line shows the basic cooling curve,
calculated using a thin quasi-stationary envelope and
the most advanced physics input, except nucleon
superfluidity; the long-dashed line is calculated using the same input
but with a traditional (thicker) blanketing envelope treated as
quasi-stationary; the dotted line is calculated using the traditional
blanketing envelope and an alternative model of radiative opacities (see
text); the dot-long-dashed line shows the result of replacement of
the opacities shown in Fig.~\ref{fig:radopac} by the simplified
opacities of \citet{PotekhinYakovlev01}; 
alternating long and short dashes demonstrate the
result of neglect of the in-medium corrections; the dot-dashed cooling
curve is calculated with allowance for nucleon superfluidity (one
selected model for each of the three types of superfluidity: neutron
singlet pairing in the crust, proton singlet and neutron triplet pairing
in the core; see text for details); the short-dashed line is calculated
for the same superfluidity model but without account of the anomalous
axial contribution to the PBF neutrino emission.
}
\label{fig:coolphys}
\end{figure}

\section{Cooling of nonmagnetized neutron stars}
\label{sect:nonmag}

Before considering the effects of superstrong magnetic fields on neutron
star cooling, let us examine the role of the microphysics updates
mentioned in Sect.~\ref{sect:physics}. The solid black curve in
Fig.~\ref{fig:coolphys} is the cooling curve of our fiducial basic
model: neutron star mass $M=1.4\,M_\odot$, BSk24 EoS, which gives
stellar radius $R=12.6$~km for this $M$, boundary
layer of mass $\Delta M=10^{-12}\,M_\odot$, which is assumed
quasi-stationary and may be partially ionized, and the physics input
described in Sect.~\ref{sect:physics}, but without baryon
superfluidity. Although the absence of baryon superfluidity seems
unrealistic, it is a convenient starting approximation for the basic
model, to which all comparisons will be made.

In the main frame of the
Figure we show evolution of the redshifted surface luminosity,
$\tilde{L}=\ee^{2{\Phi/c^2}}{L}_R$.
Errorbars and arrows show
observational estimates of the ages and thermal luminosities
of 44 neutron stars with confirmed thermal emission. The
first 41 sources are taken from Table~3 of
\citet{PotekhinPP15}, and are numbered according to their
order in that table, which was derived from the catalog of
\citet{Vigano_ea13} with the addition of one source, PSR
J1741$-$2054 (object number 13,
\citealp{Karpova_ea14,Marelli_ea14}). For the thermal
luminosity of the neutron star CXO J232327.9+584842 in the
supernova remnant Cassiopeia A (object number 4, often
dubbed Cas A NS) we have adopted improved observational data
on $\tilde{L}$ \citep{Posselt_ea13}. We have supplemented
this catalog with three neutron stars with recently measured
thermal fluxes: two neutron stars with carbon envelopes
(sources 42 and 43, \citealp{Klochkov_ea15,Klochkov_ea16}),
and one pulsar (source 44,
\citealp{Danilenko_ea15,Karpova_ea17}; in this case, we
adopt the interpretation of the thermal spectrum with the
model of a partially ionized magnetized hydrogen atmosphere
\textsc{nsmax}, \citealp{HoPC08}). Objects 25 through 41
(red symbols) are magnetars.

The horizontal errorbars show estimates of the lower and upper
bounds on the kinematic age of the star, determined from observations of
the related supernova remnants. In the cases where only an estimate
without errors is available, we replace the errorbar by a short
double-sided arrow.
In the cases where the kinematic age is unavailable, we use an estimate
of the characteristic age, determined from the stellar spin period and
its derivative in the canonical model of the rotating magnetic dipole in
vacuum. The characteristic ages are measured very accurately, but they
are rather poor estimators of a true age, therefore we plot longer
horizontal arrows in these cases. 

In the inset we show evolution of the nonredshifted effective
temperature $\Teff$ in a restricted time interval. Here, the points
represent estimates of $\Teff$ for the Cas A NS from Table~I of
\cite{Ho_ea15}. The difference between the vertical
position of these points in the
inset and the errorbar 4 in the main frame is caused by the use of a
different neutron star model, which gives a larger redshift and higher
$\Teff$ (see \citealp{Elshamouty_ea13}).

\begin{figure}
\centering
\includegraphics[width=\columnwidth]{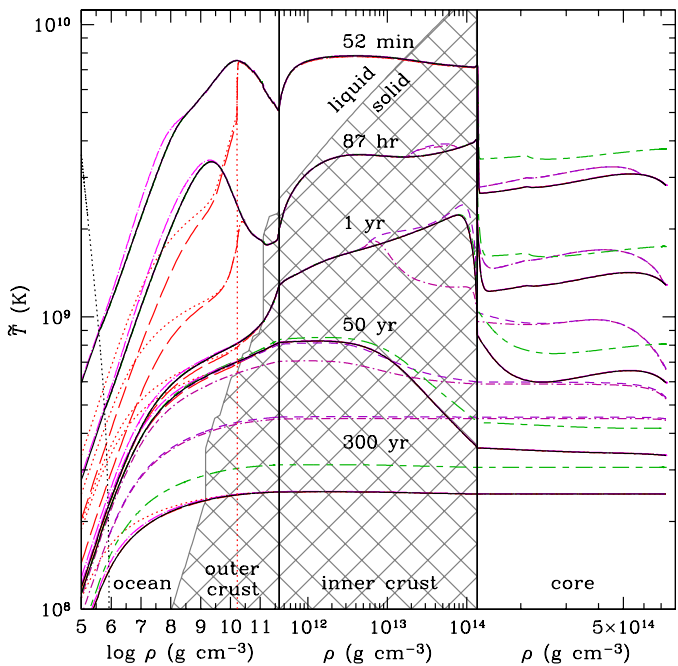}
\caption{Redshifted temperature as a function of mass density inside a
neutron star with $M=1.4\,M_\odot$ at different ages (marked near the
curves) according to different theoretical models. Line styles
correspond to theoretical models in the same way as in
Fig.~\ref{fig:coolphys}. The three parts of the Figure show the thermal
structure of the envelopes comprising the ocean and the outer
crust ((left), the inner crust (middle), and the core (right)
at different density scales. In the gray-cross-hatched 
$\rho-\tilde{T}$ domain, the nuclei are arranged in a
crystalline Coulomb lattice. The nearly vertical dotted lines in the
left part of the Figure show the position of the outer boundary for the
cooling code, beyond which (to the left in the figure) the heat
transport problem is solved using quasi-stationary approximation: the
left (black) and right (red) lines delimit a quasi-stationary envelope
of mass $\Delta M=10^{-12}\,M_\odot$ and $\Delta M=10^{-6}\,M_\odot$,
respectively.
}
\label{fig:structphys}
\end{figure}

\subsection{Thickness of the quasi-stationary envelope}
\label{sect:thick}

The cooling curves marked ``a'' and ``b'' in
Fig.~\ref{fig:coolphys} are obtained for a model, which differs from the
basic model by the thickness of the envelope that is treated in
the quasi-stationary approximation, $\Delta
M=10^{-6}\,M_\odot$ (in addition, curve ``b'' differs by the
radiative opacity approximation, see Sect.~\ref{sect:opac}). For the
given $M$ and $R$, the mass density at the bottom of this
``thick'' envelope is
$\rho\approx1.7\times10^{10}$ \gcc, which is close to the
traditional $\rho_\mathrm{b}$ value. Comparison of the curve
``a'' with the basic cooling curve shows that the simulations with
the traditional blanketing envelope are quite accurate on the timescales
$t\gtrsim1$~yr, but not on shorter timescales. The reason for this can
be understood from inspection of Fig.~\ref{fig:structphys}, which shows
temperature as function of density. At early ages
the temperature profile in the thick envelope obtained in
the quasi-stationary approximation (lines ``a'' and ``b'') strongly
differs from the profile calculated with the full account of
time dependence in the thermal evolution equation
(\ref{heat_diffusion}) (the solid line, ``basic model'').
The solution of this equation in the quasi-stationary
approximation is equivalent to setting
$c_P=0$. Thus we see that the accurate
evaluation of $c_P$ is important in the outer crust and
ocean of a neutron star at relatively short
timescales $t\lesssim1$ yr.

\subsection{Opacities}
\label{sect:opac}

The influence of the model for radiative opacities on the cooling curves
can be seen from comparison of the basic (solid) curve in
Fig.~\ref{fig:coolphys} with the dot-long-dashed curve 
(marked ``opac. PY'01''), which is calculated with simplified
opacities following a model that does not include the plasma cutoff,
electron-positron pairs, Compton effect, or electron degeneracy 
\citep{PotekhinYakovlev01}. The difference between the two cooling
curves is noticeable only at high $\tilde{L}$ (the difference can
exceed 10\%, only if $\tilde{L} \gtrsim 10^{36.5}$ erg s$^{-1}$; in the
latter case, the effect is mainly due to the importance of Compton
scattering at high temperatures).

In the atmosphere, ionization can be incomplete, which is  especially
important in the case of heavy elements or strong magnetic fields. For
the nonmagnetic iron atmosphere, we use the \textsc{opal} radiative
opacities. There can be an ambiguity in connecting this opacity to the
radiative opacity calculated in the model of a fully ionized plasma,
shown in Fig.~\ref{fig:radopac}. In the case of the thick blanketing
envelope, there are two curves in Fig.~\ref{fig:coolphys}, long-dashed
and dotted ones, corresponding to different prescriptions for the
connection. The first one (curve ``a'') is obtained by using the
\textsc{opal} opacities in the density-temperature range where they are
tabulated and replacing them by the fully ionized plasma model beyond
the tables. The second one (curve ``b'') is obtained with extrapolation
of the tabular data to the bottom of the blanketing envelope, using the
Kramers opacity law. The thermal structure of the envelopes with using
these two prescriptions is illustrated in Fig.~\ref{fig:structphys} by
the red lines of the same styles (dotted and long-dashed) as
corresponding lines ``a''
and ``b'' in Fig.~\ref{fig:coolphys}. We see that these two ways of handling
the radiative opacity data can result in noticeably different
cooling curves, the difference being especially pronounced at
$t\lesssim100$~yr, because at early ages the radiative transport
dominates in a large part of the envelope due to the high temperatures.

\subsection{Anomalous axial PBF contribution}

To check the importance of the recent advances in the treatment of the
Cooper PBF neutrino emission mechanism, we first include the
superfluidity with the accurate treatment of this emission mechanism
according to \citet{Leinson10} (the dot-dashed lines in
Figs.~\ref{fig:coolphys} and~\ref{fig:structphys}), and then switch off
the anomalous contribution into the axial channel of weak interactions
(the short-dashed lines). For illustration, we have chosen the
superfluidity type SF081326. Here and hereafter, the first (08), second
(13), and third (26) pair of digits compose the entry number in
Table~II of \citet{Ho_ea15}. Thus, SF081326 stands for the superfluid
gap model SFB for the neutron singlet superfluidity \citep{SchwenkFB03},
CCDK for the proton singlet superfluidity
\citep{Chen_ea93,Elgaroy_ea96}, and TToa for the neutron triplet
superfluidity \citep{TakatsukaTamagaki04}. This combination has been
selected by \citet{Ho_ea15} as the best fit to the apparent surprisingly
rapid decline of luminosity of the Cas A NS, first reported by
\citet{HeinkeHo10}. In the inset of Fig.~\ref{fig:coolphys} we can see
that the short-dashed curve $\Teff(t)$ is indeed close to the
plotted Cas A NS data in the common approximation neglecting the
anomalous axial contribution, but it is not the case if  the PBF
emission is calculated accurately. This conclusion confirms the
statement by \citet{Leinson16} about the importance of the latter
contribution.

Let us note that \citet{Shternin_ea11} were the first who
encountered the impossibility of fitting the apparent fast decline of
the Cas A NS luminosity with the results of \citet{Leinson10}. In order
to obtain an acceptable fit to this decline, they (as later
\citealt{Elshamouty_ea13}) arbitrarily increased the PBF reaction rate
by changing the coefficient 0.19, mentioned in
Sect.~\ref{sect:neutrino}, to 0.4, which does not follow from any
theory. This artificial enhancement of the PBF rate is however
unnecessary, since the fast cooling of Cas A NS has been put in doubt by
\citet{Posselt_ea13}, who suggested possible alternative explanations to
the observational data.

\begin{figure}
\centering
\includegraphics[width=\columnwidth]{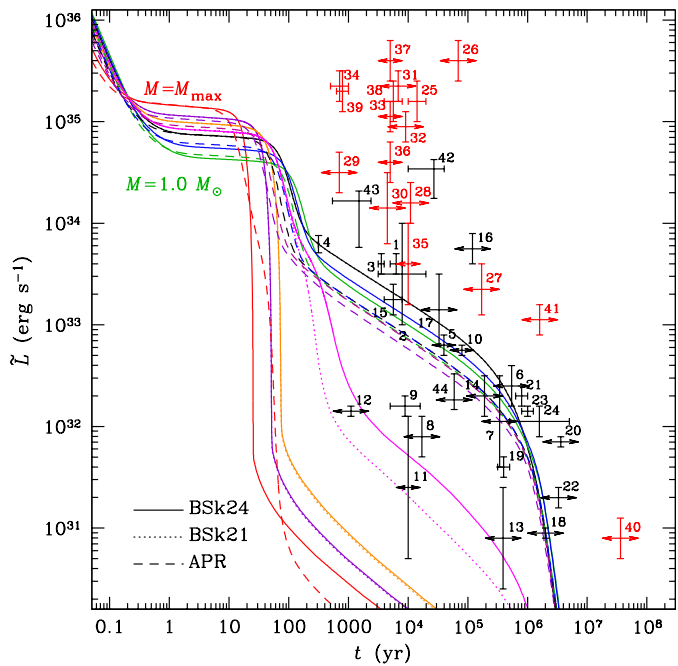}
\caption{The cooling curves for neutron stars with $M=1.0$ (green),
1.2 (blue), 1.4 (black), 1.6 (magenta), 1.8 (orange), $2.0\,M_\odot$
(violet), and the maximum mass $M_\mathrm{max}$ (red lines),
calculated using the same basic theoretical model as in
Fig.~\ref{fig:coolphys}, for the EoS models BSk24/21 (solid/dotted lines,
$M_\mathrm{max}=2.28\,M_\odot$, $M_\mathrm{DU}=1.596/1.587\,M_\odot$)
and APR+NV (dashed lines, $M_\mathrm{max}=2.21\,M_\odot$,
$M_\mathrm{DU}=2.01\,M_\odot$) compared with observations (the same
symbols as in Fig.~\ref{fig:coolphys}).
}
\label{fig:coolmass}
\end{figure}

\subsection{In-medium effects}

The importance of the in-medium effects can be seen from comparison of
the corresponding cooling curves in Fig.~\ref{fig:coolphys} and
temperature profiles in Fig.~\ref{fig:structphys}. Simulations without
account of these effects substantially overestimate luminosities and
effective temperatures of middle-aged neutron stars.
We have checked that
this effect is caused mainly by the in-medium enhancement of the
modified Urca reactions \citep{ShterninBaldo17}. The in-medium effects
on thermal conductivities \citep{BaikoHY01,ShterninBH13} turn out to be
unimportant.

\subsection{Equation of state and direct Urca processes}
\label{sect:EoSDU}

\begin{figure}
\centering
\includegraphics[width=\columnwidth]{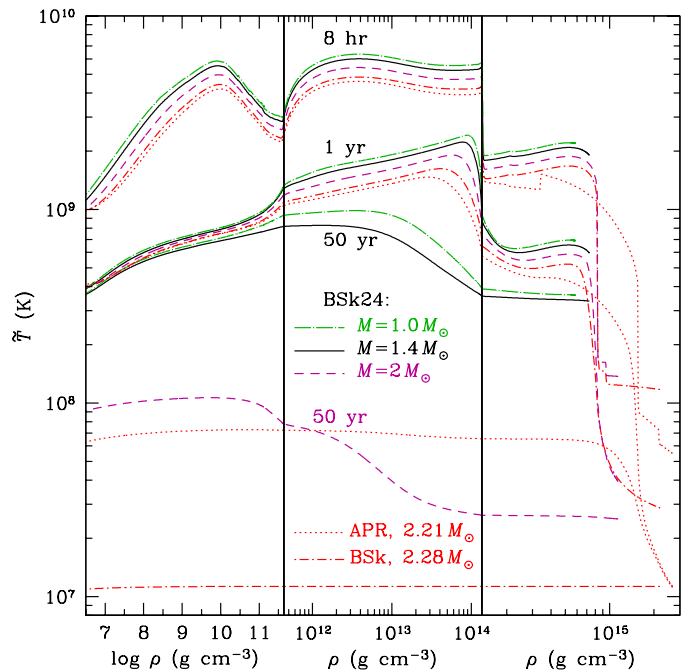}
\caption{Redshifted temperature as function of mass density inside
neutron stars with masses $M=1.0$ (dot-long-dash lines), 1.4 (solid
lines), 2.0 (short-dashed lines), and $M_\mathrm{max}=2.28\,M_\odot$
(dot-short-dash lines) for the unified EoS model BSk24, and with mass
$M_\mathrm{max}=2.21\,M_\odot$ for the APR+NV EoS model (dotted lines)
at ages of $10^{-3}$, 1, and 50 years. The vertical lines separate
three density regions: the ocean and outer crust, the inner crust, and the core,
displayed using different density scales.
}
\label{fig:structmass}
\end{figure}

Figures \ref{fig:coolmass} and \ref{fig:structmass} illustrate the
effects of varying neutron star mass and equation of state on the
cooling curves and thermal structure of neutron stars. In
Fig.~\ref{fig:coolmass} we show a sequence of cooling curves for stellar
masses $M$ from $1.0\,M_\odot$ to $2.0\,M_\odot$ and the maximum mass
$M_\mathrm{max}$, according to two EoSs, BSk24
($M_\mathrm{max}=2.28\,M_\odot$) and APR
($M_\mathrm{max}=2.21\,M_\odot$).

The cooling curves are close to the
basic curve until $M$ exceeds the direct Urca threshold $M_\mathrm{DU}$,
which is equal to $1.596\,M_\odot$ for BSk24 and to $2.01\,M_\odot$ for
APR. The difference of $M_\mathrm{DU}$ for different EoSs is caused by
different density-dependence of the electron fraction, which is obtained
along with the thermodynamic functions while calculating the EoS by the
free energy minimization. For $M>M_\mathrm{DU}$, a powerful cooling
occurs in the central part of the core.

We have also performed analogous calculations for the EoS model BSk21,
but found that the results are almost indistinguishable from those
obtained with BSk24, except for values of $M$ close to $M_\mathrm{DU}$.
In Fig.~\ref{fig:coolmass}, the difference is noticeable only for
$M=1.6\,M_\odot$ (the dotted curve). This is because the sensitivity of
the cooling to the mass of the central region where the direct Urca
processes operate. For BSk21, $M_\mathrm{DU}=1.587\,M_\odot$, the
threshold at $M=1.6\,M_\odot$ is exceeded by
$M-M_\mathrm{DU}=0.013\,M_\odot$, compared to $0.004\,M_\odot$ for
BSk24. A change of $M$ by $-0.009\,M_\odot$ brings the dotted cooling
curve in Fig.~\ref{fig:coolmass} to the solid one. Such a small
difference in $M$ seems to be practically insignificant.

Figure~\ref{fig:structmass}
shows temperature profiles for several
masses and ages. 
We see that some of the red and magenta lines go down towards the
right edge of the Figure, indicating temperature decrease with
increasing density towards the center of the most massive stars. At
early ages ($t\lesssim1$ yr), this temperature decrease exceeds an
order of magnitude for the models with $M=2\,M_\odot$ and
$M=M_\mathrm{max}$ using the BSk24 EoS and for the model with
$M=M_\mathrm{max}$ using the APR EoS. Thus the central region
of the massive neutron star works as a cooler. 
As seen from Fig.~\ref{fig:structmass}, it is effective until
the temperature falls to $\sim10^7$~K. Afterwards the neutrino cooling
dyes away and the temperature profile flattens out.

\begin{figure}
\centering
\includegraphics[width=\columnwidth]{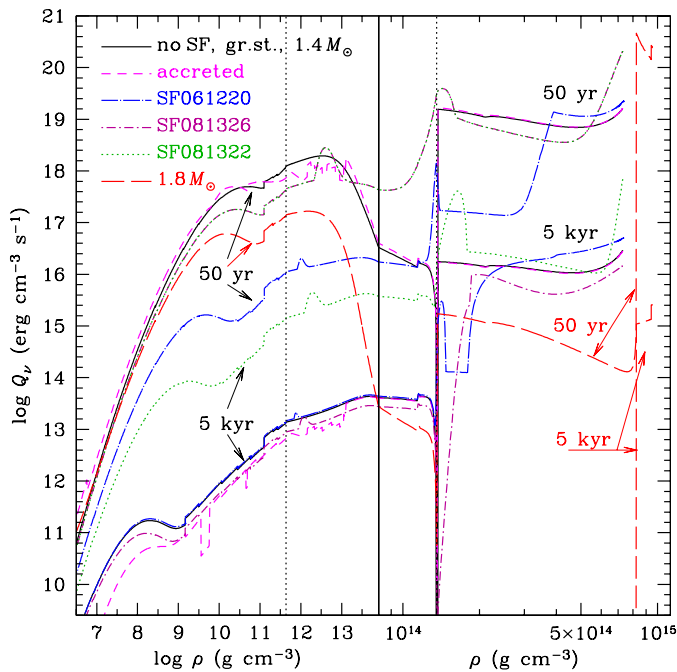}
\caption{Neutrino emission power density as function of mass density at
ages 50 yr (upper curves) and 5 kyr (lower curves of each type) for
different neutron star models: the basic model (EoS BSk24, ground state
composition, $M=1.4\,M_\odot$, no superfluidity; solid lines), the
neutron star with fully accreted envelope (short-dashed lines), three
different combinations of the three types of superfluidity
(dot-long-dash, dot-short-dash, and dotted curves), and a model without
superfluidity but with the direct Urca processes due to the higher mass
$M=1.8\,M_\odot$ (long-dashed lines).
The density scale is different in the left and right parts of the
Figure, separated by the vertical solid line. The vertical dotted lines
mark the outer and inner boundaries of the inner crust.
}
\label{fig:emission}
\end{figure}

The total neutrino emissivities for several neutron star
models and two ages are shown in Fig.~\ref{fig:emission}. We see that
for $M=1.8\,M_\odot>M_\mathrm{DU}$(BSk24) the emission rate at
$\rho>8.25\times10^{14}$ \gcc{} is enhanced by several orders of
magnitude compared to the rates at lower densities.

\subsection{Superfluidity}

\begin{figure}
\centering
\includegraphics[width=\columnwidth]{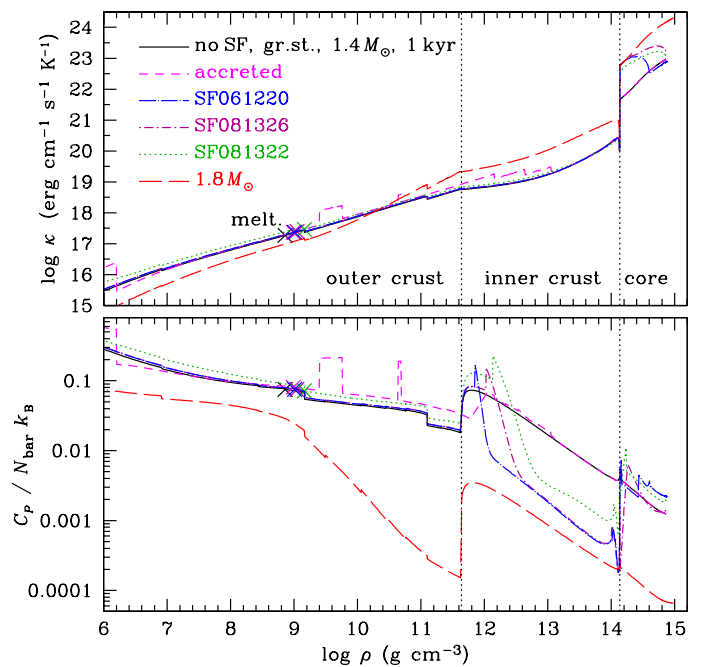}
\caption{Thermal conductivity (upper panel) and heat capacity per baryon
in units of $\kB$ (lower panel) as functions of mass density at age
$t=1$ kyr for the same neutron star models as in Fig.~\ref{fig:emission}.
The crosses mark the melting points for the models with
$M=1.4\,M_\odot$: to the left of them, the Coulomb plasma forms a liquid
ocean. For the $M=1.8\,M_\odot$ model, the crust does not melt in the
displayed density range, because it is relatively cold at
this age (cf.\ Fig.~\ref{fig:coolmass}). The vertical dotted lines
mark the outer and inner boundaries of the inner crust.
}
\label{fig:cpkappa}
\end{figure}

It is well known that superfluid energy gaps for different
types of nucleon pairing have an important influence on the neutron star-cooling curves. When temperature falls substantially below $\Tcrit$ for
some kind of the baryons, the superfluidity reduces their heat capacity
\citep{LevenfishYakovlev94,YakovlevLS99}. The baryon superfluidity may
also either reduce or increase the conductivity in the core
\citep{BaikoHY01}. These effects are seen respectively in the lower and
upper panels of Fig.~\ref{fig:cpkappa}. 

The baryon superfluidity affects the neutrino emissivity in different
ways \citep[][and references therein]{YKGH}. 
Neutrino emission by the modified and direct Urca processes and by
baryon bremsstrahlung gets strongly suppressed at $T\ll\Tcrit$. However,
when $T$ is close to $\Tcrit$, the total neutrino emission is greatly
enhanced due to the PBF mechanism. Since $\Tcrit$ for each type of
superfluidity depends on baryon density, the overall picture is
complicated, as we see in Fig.~\ref{fig:emission}.

\begin{figure}
\centering
\includegraphics[width=\columnwidth]{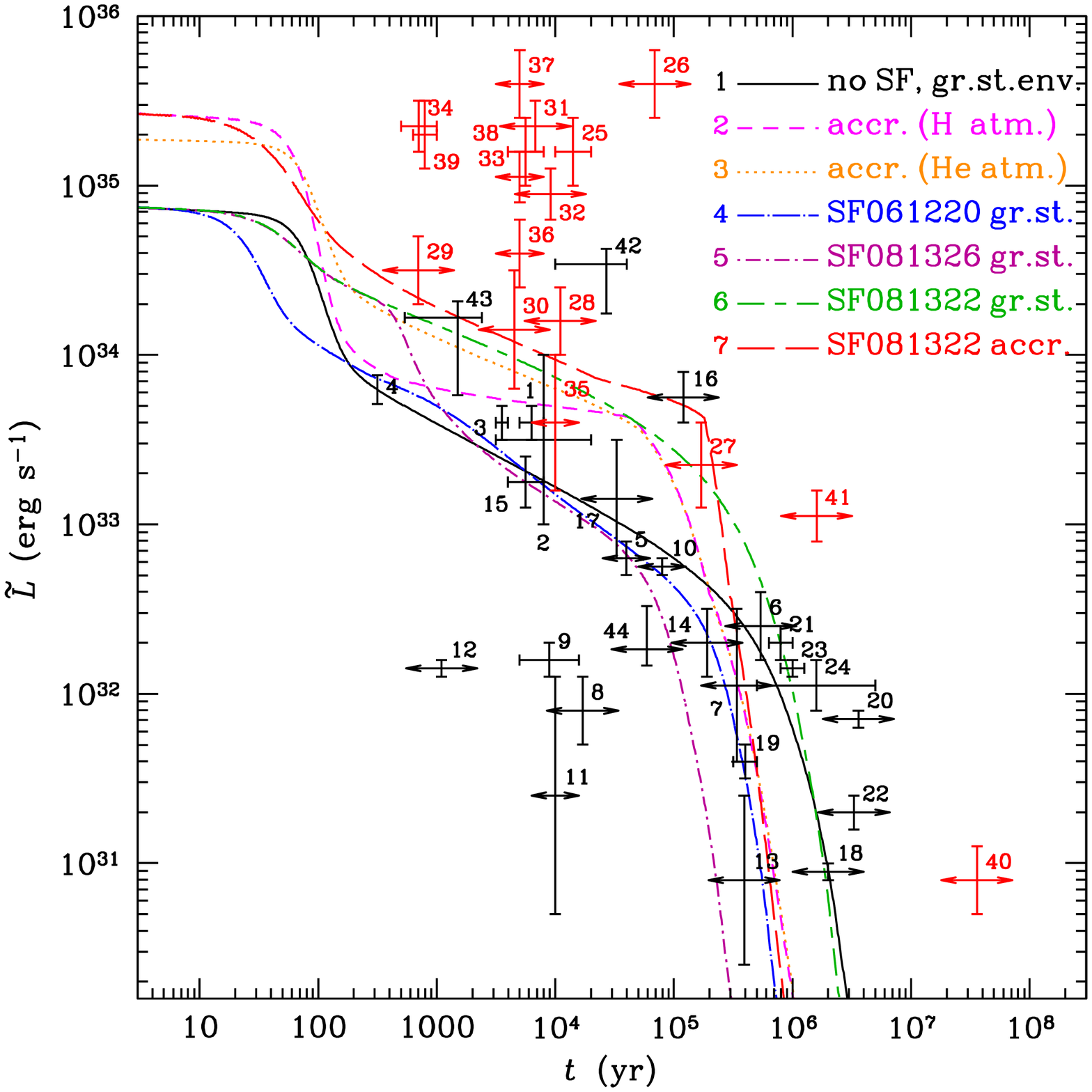}
\caption{Cooling curves for a neutron star with $M=1.4\,M_\odot$
according to the BSk24 EoS model with iron (lines 1, 4, 5, 6) or
accreted envelopes (lines 2, 3, 7),  iron (lines 1, 4, 5, 6), hydrogen
(lines 2, 7), or helium atmosphere (line 3), without nucleon
superfluidity (lines 1\,--\,3) and with different models of nucleon
superfluidity (lines 4\,--\,7; see text for explanation of the
SF-notations), compared with observations (the same symbols as in
Fig.~\ref{fig:coolphys}).
}
\label{fig:coolaccr}
\end{figure}

\begin{figure}
\centering
\includegraphics[width=\columnwidth]{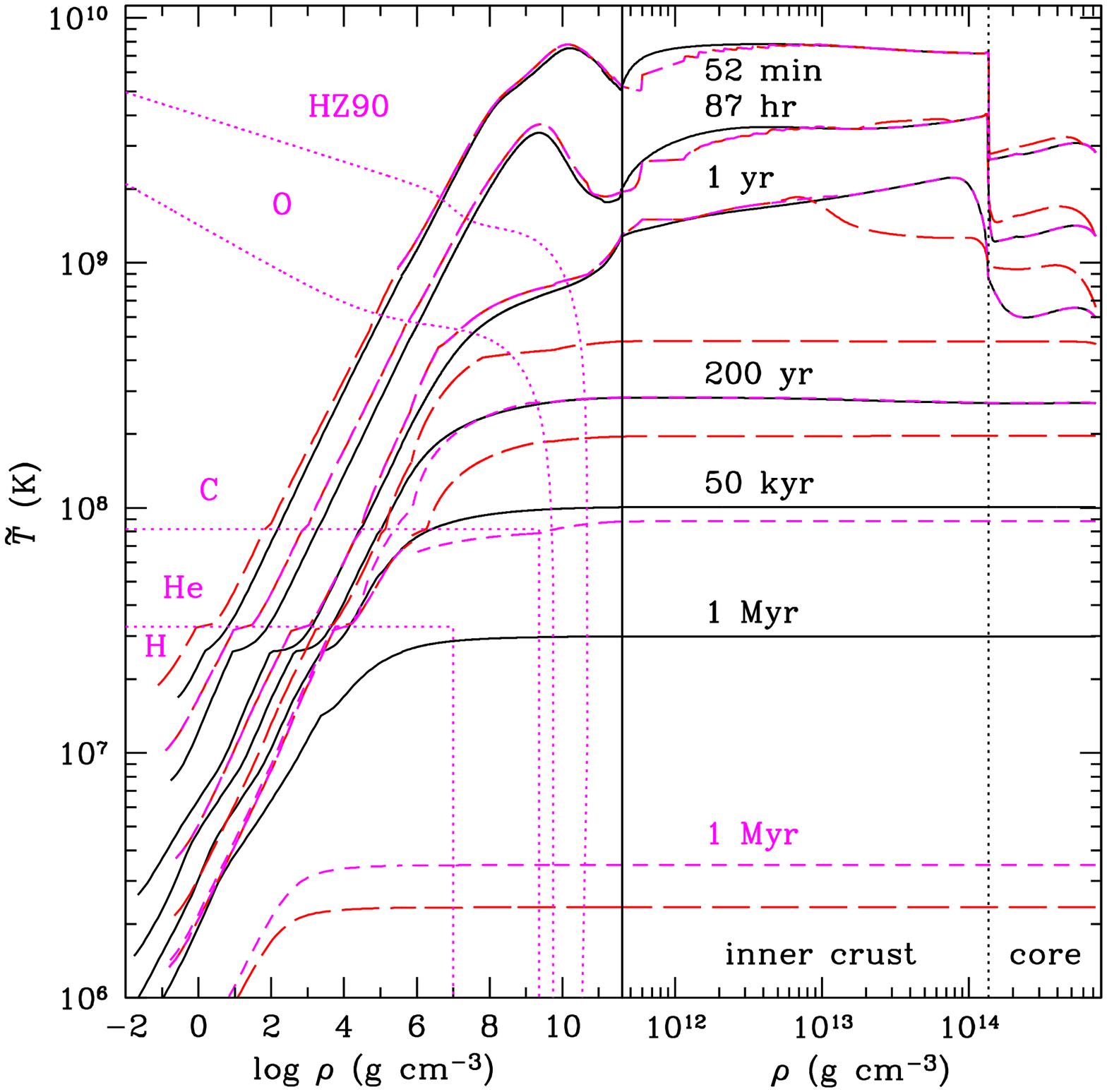}
\caption{Redshifted temperature as a function of mass density inside a
neutron star for three of the seven different models considered in
Fig.~\ref{fig:coolaccr} (namely, models 1, 2, and 7), plotted using the same line styles, at
different ages marked near the curves. The two parts of the Figure,
separated by the
vertical solid line, show
the regions before and after neutron drip at different density scales.
The dotted lines in the left half of the Figure show the adopted
density-temperature domains for different chemical elements (H, $^4$He,
$^{12}$C, $^{16}$O) in the accreted envelope. To the right of the last
of these lines, the composition of the accreted crust is adopted from
\citet{HaenselZdunik90}. For the nonaccreted crust, we use the
ground-state composition of matter according to the BSk24 model. The
vertical dotted line in the right half of the Figure separates the crust
and the core.
}
\label{fig:structaccr}
\end{figure}

As a result, some parts of a neutron star may have higher- while other
parts lower temperature than it would be for a neutron star without
superfluidity. We see this in Figs.~\ref{fig:structphys} and
\ref{fig:structaccr}, which show temperature profiles in superfluid and
nonsuperfluid neutron stars of different ages. The cooling can be
accelerated or decelerated, depending on the interplay of the
superfluidity-related
mechanisms, as we see in Fig.~\ref{fig:coolaccr} where the effects of
three different models of superfluidity are demonstrated.

The model SF081326, which is also used in Figs.~\ref{fig:coolphys} and
\ref{fig:structphys}, is characterized by a relatively weak neutron
singlet-pairing superfluidity SFB in the crust (maximum of the critical
temperature $\max\Tcrit\sim5\times10^{9}$~K), rather strong proton singlet
superfluidity CCDK in the core ($\max\Tcrit\approx7\times10^9$~K), and
moderate neutron triplet superfluidity TToa in the core
($\max\Tcrit\approx6\times10^8$~K). In Fig.~\ref{fig:coolaccr} we
reproduce the corresponding cooling curve (line 4) and compare it
with the curves calculated using two different combinations of the
superfluidity models (lines 5 and 6).
Retaining the same singlet-type superfluidities but replacing the
moderate neutron triplet model TToa by the weak neutron triplet
superfluidity model EEHOr with $\max\Tcrit<2\times10^8$~K
\citep{Elgaroy_ea96b}, we  effectively suppress the PBF neutrino
emission at moderate ages, which helps to keep a neutron star relatively
hot for a long time (curves 6 and 7 in Fig.~\ref{fig:coolaccr}). This
model is sufficient for explanation of several sources showing luminosities above those
predicted by the basic cooling curve; nevertheless, there remain many still
hotter neutron stars, which require another explanation.

As another example, we choose model combination SF061220: the strong
neutron singlet superfluidity MSH ($\max\Tcrit>10^{10}$~K,
\citealp{MargueronSH08,Gandolfi_ea09}), moderate proton singlet
superfluidity BS ($\max\Tcrit\sim5\times10^9$~K,
\citealp{BaldoSchulze07}), and moderate neutron triplet superfluidity
BEEHS ($\max\Tcrit\sim4\times10^8$~K, \citealp{Baldo_ea98}). In this
case, cooling is accelerated at early and late ages
(line 4 in Fig.~\ref{fig:coolaccr}).

\subsection{Accreted envelopes}

The neutron star envelopes are more transparent to the heat flux if
they are composed of light chemical elements. Previously we studied this
effect in the quasi-stationary approximation for nonmagnetic
\citep{PCY97} and strongly magnetized envelopes \citep{Potekhin_ea03}.
The main effect is related to the $Z$-dependence of the electron-ion 
collision frequencies. The higher the ion charge number $Z$, the
larger the collision frequency and the lower the conductivity.
Another effect concerns radiative opacities in the photosphere, which
are also smaller for light elements than for iron.

Here we
more accurately simulate the cooling of neutron stars with accreted envelopes by relaxing the quasi-stationary approximation. The
cooling curves for nonsuperfluid (lines 1\,--\,3) and superfluid
(lines 4\,--\,7) neutron stars with nonaccreted (lines
1, 4, 5, 6) and accreted (lines 2, 3, 7) envelopes are compared in
Fig.~\ref{fig:coolaccr} and their thermal structures are shown in
Fig.~\ref{fig:structaccr}. 

For the accreted envelopes, we use
either hydrogen (lines 2 and 7) or helium (line 3) atmosphere
models.
An interesting effect, first noticed by
\citet{BeznogovPY16}, is that the replacement of H by $^4$He increases
the photon luminosity. This effect is caused by the different mass to
charge ratio of H, which results in the discontinuity of $\rho$ and is
related to the lower opacities of He at high $T$and fixed $\rho$,
revealed in both \textsc{op} and \textsc{opal} tables.
These discontinuities are seen
in the left part of Fig.~\ref{fig:structaccr} at the H/He
interface,
which is set at $T=4\times10^7$~K ($\tilde{T}\approx3.3\times10^7$~K),
for the accreted-envelope models (dashed lines). On the other hand, the
kinks at the temperature profiles for nonaccreted envelopes (solid lines
in Fig.~\ref{fig:structaccr}) at $\tilde{T}\lesssim3\times10^7$~K are
caused by the switch to the fully ionized plasma model for radiative
opacities at the boundary of the \textsc{opal} tables for iron (see
Sect.~\ref{sect:opac}).

 The composition of the accreted crust at larger densities, beyond the
 C and O layers,
is produced by a sequence of nuclear transformations during
accretion \citep{HaenselZdunik90}. This leads to
different thermal conductivities and heat capacities, as seen in
Fig.~\ref{fig:cpkappa}, and to different rates of
neutrino bremsstrahlung, as seen in Fig.~\ref{fig:emission}. The latter differences, however, are
not sufficiently strong to noticeably affect the cooling curves.

Figure~\ref{fig:coolaccr} shows that the largest theoretical
luminosities are obtained by the use of the model with strong proton
singlet and weak neutron triplet superfluidity (SF081322) and the
accreted He envelope.

\subsection{Other effects}

We have additionally tested a number of other updates to the neutron
star microphysics, which however proved to be unimportant. Two examples,
the in-medium corrections to thermal conductivities and the upgrade of
the BSk21 to the BSk24 EoS, have been mentioned above. Another example
is latent heat of the crust. As a neutron star cools down, its frozen
crust grows, gradually replacing the ocean. We have included the latent
heat released during this process as a heat source $\tilde{H}$ in
\req{heat_diffusion}, but found the cooling curves unchanged. This
should have been anticipated, because the crust contains only a small
fraction (less than 2\%) of the total mass $M$ of a neutron star. 

The replacement of the fit of \citet{YKGH} to bremsstrahlung neutrino
energy losses in the crust by a more detailed fit of
\citet{OfengeimKY14} also does not affect the cooling curves,
even in the case of accreted crust

We have also checked that the replacement of the fit of \citet{YKGH} to
the plasmon decay rate by the more accurate fit of
\citet{KantorGusakov07} does not have a noticeable effect on the cooling
curves (this statement pertains to neutron stars, but not to white
dwarfs, where the plasmon decay is an important channel of energy losses
at early ages).

\section{Cooling of magnetars}
\label{sect:magnetars}

The effects of quantizing magnetic fields on the thermal structure of
neutron-star envelopes were first studied by \citet{Hernquist85} and
somewhat later by \citet{VanRiper88} and \citet{Schaaf90}.
\citet{VanRiper88} considered a neutron star with a constant radial
magnetic field. In this model, the quantum enhancement of longitudinal
electron conductivity $\kappa_\|$ at $\rho\sim\rho_B$
(where $\rho_B$ is given by \req{rho_B})
results in an overall enhancement of the
neutron-star photon luminosity at a fixed internal temperature. However,
\citet{ShibanovYakovlev96} showed that, for the dipole field
distribution, the effects of suppression of the heat conduction across
the field near the magnetic equator compensate or even overpower the
effect of the conductivity increase near the   normal direction of the
field lines. This conclusion was also shown to be valid  for small-scale
fields at $B\lesssim10^{14}$~G \citep{PotekhinUC05}. However, in
superstrong fields  $(B\gtrsim10^{14}$~G) the quantum enhancement of the
longitudinal conductivity and the corresponding increase of the stellar
photon luminosity are more important. This enhancement overpowers the decrease of the
luminosity in the regions of almost tangential field lines, so that
overall photon luminosity increases \citep{Kaminker_ea09,Kaminker_ea14}.
This may not be the case in the configurations where the field is nearly
tangential over a significant portion of the stellar surface as, for example,
in the case of a superstrong toroidal field
\citep{PerezAzorinMP06,PageGK07}. We do not study the latter possibility
in the present paper, assuming that the superstrong magnetar field is
highly tangled.

\begin{figure}
\centering 
\includegraphics[width=\columnwidth]{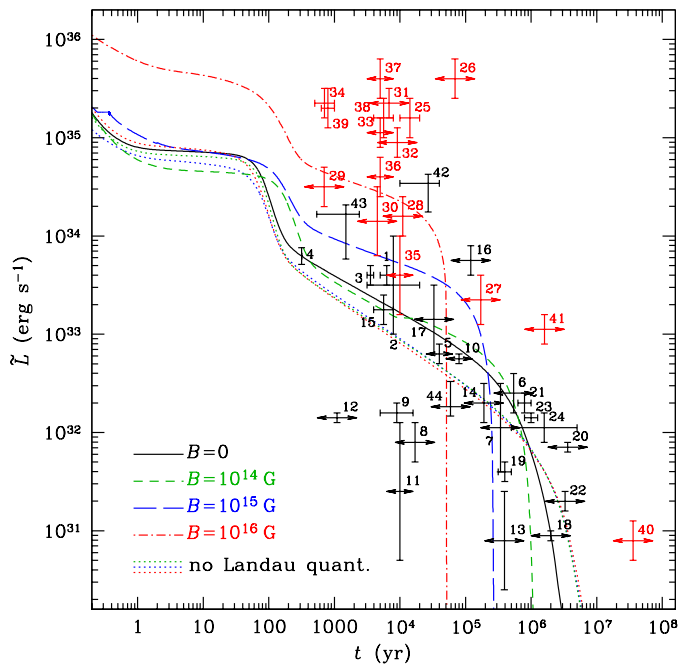}
\caption{Cooling curves for neutron stars with $M=1.4\,M_\odot$
according to the BSk24 EoS model with $B=0$ (solid line), $10^{14}$~G
(short-dashed line), $10^{15}$~G (long-dashed line), and $10^{16}$~G
(dot-dashed line). The three dotted lines result from simulations with
the same three strong fields, but treated classically.
}
\label{fig:coolquant} 
\end{figure}

Figure~\ref{fig:coolquant} shows cooling curves of neutron stars with
$B=10^{14}$~G, $10^{15}$~G, and $10^{16}$~G, compared with the cooling
of a nonmagnetic neutron star. 
We see the strong enhancement of the
luminosity for middle-aged neutron stars with superstrong magnetic
fields. As a consequence of rapid energy loss,  the stored thermal
energy is spent more rapidly, and the luminous lifetime of the magnetars
becomes shorter. We intentionally leave aside additional heating by the
magnetic energy consumption (e.g., by one of the mechanisms
considered by \citealp{BeloborodovLi16}), which may become the subject of a
separate study. Here we aim at revealing the luminosity increase that can
be provided by the effects of the quantizing magnetic fields on the
crust of the star. 

For comparison, we plot the cooling curves obtained using classical
treatment of the magnetic fields (i.e., without allowance for the Landau
quantization). At the middle ages of the stars, the latter curves pass
below the nonmagnetic curve, which is explained by the magnetic
suppression of the electron heat transport across the field lines.

\begin{figure}
\centering
\includegraphics[width=\columnwidth]{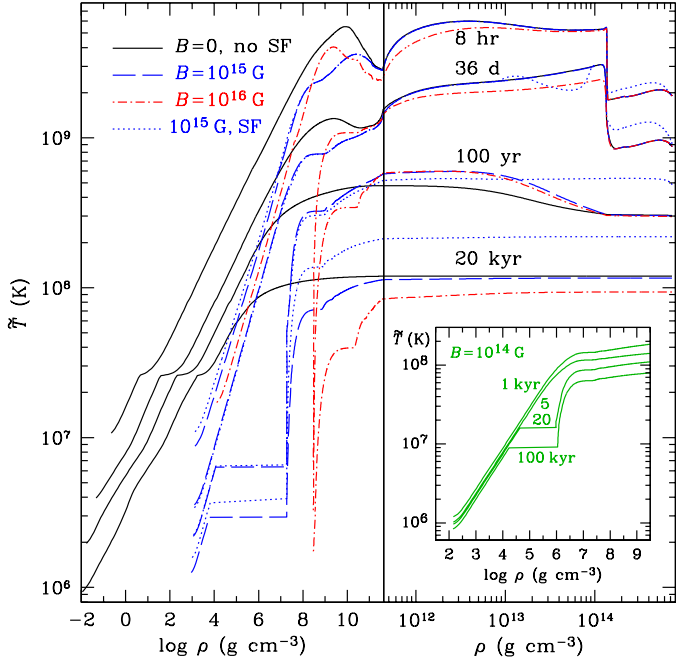}
\caption{Redshifted temperature as a function of mass density inside
neutron stars with $B=0$ (solid lines), $10^{15}$~G (long-dashed and
dotted lines), and $10^{16}$~G (dot-dashed lines) at four ages $t$ from
8 hours to $2\times10^4$ years, marked near the curves. The dotted line
illustrates the model with superfluidity SF081322; the other lines are
calculated without superfluidity.
The two parts of the Figure,  separated by the vertical solid line,
show the regions before and after neutron drip at different density
scales.
Inset: temperature profiles in the
envelopes of a neutron star with $B=10^{14}$~G at ages $t=1$, 5,
20, and 100 kyr.
}
\label{fig:structmag}
\end{figure}

\begin{figure}
\centering 
\includegraphics[width=\columnwidth]{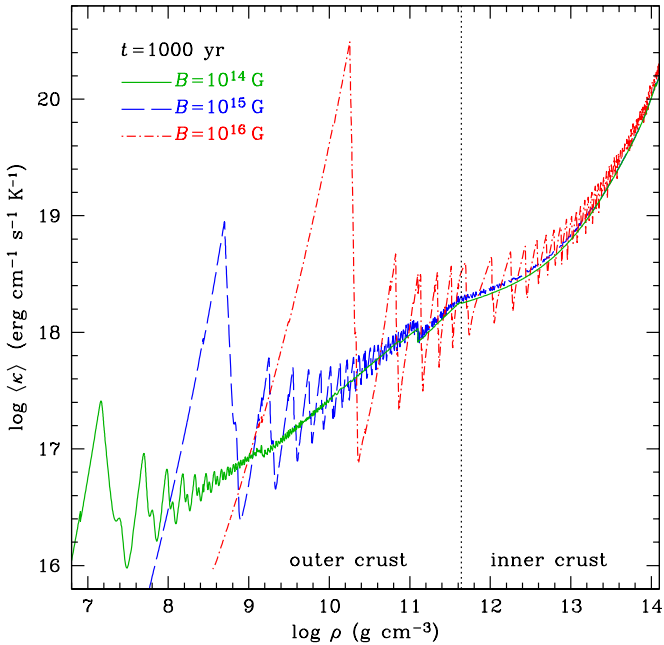}
\caption{Effective thermal conductivity as a function of density in the
ocean and crust of a kyr-old magnetar with magnetic fields
$B=10^{14}$~G, $10^{15}$~G, and $10^{16}$~G.
The vertical dotted line marks the neutron drip density.
}
\label{fig:kappamag} 
\end{figure}

Figure \ref{fig:structmag} shows thermal structure of magnetars at
different ages. For comparison, we reproduce the thermal structure of a
nonmagnetized neutron star with iron envelopes from
Sect.~\ref{sect:nonmag}. The temperature profiles of the strongly
magnetized neutron stars reveal series of kinks, which are related to
the magnetic quantization. They are located near the points where
degenerate electrons fill consecutively excited Landau levels at
densities $\rho\gtrsim\rho_B$. At these densities, the average
conductivities have peaks and dips, as shown in Fig.~\ref{fig:kappamag}.

Only part of the enhancement of luminosity in
Fig.~\ref{fig:coolquant} is due to the above-mentioned increase of
conductivity at $\rho\sim\rho_B$. Equally important is the magnetic
condensation, the phenomenon predicted by \citet{Ruderman71} and studied
by \citet{LaiSalpeter97} and \citet{MedinLai06,MedinLai07}. In a
nonmagnetized dense nonideal electron-ion plasma, the pressure of
degenerate electrons overpowers the negative electrostatic pressure at
high densities. In the superstrong fields of magnetars, however, the
Fermi temperature is so strongly reduced at $\rho\ll\rho_B$, that the
electrons become nondegenerate, and hydrostatic instability develops,
leading to formation of a condensed surface at high density. The
radiation escapes directly from this hot surface, without diffusion
through a gaseous atmosphere. Such a neutron star is said to be ``naked''
\citep{TurollaZD04}. 

The process of formation of the naked neutron star is elucidated by
Fig.~\ref{fig:structmag}. 
At early times, when a magnetar is sufficiently hot, its thermal
structure is smooth because it possesses a thick atmosphere. As the
magnetar cools down, a sharp condensed surface appears. Nevertheless, in
the case of $B=10^{15}$~G, we see that the neutron star
retains  a finite atmosphere with densities $\rho\gtrsim 10^3$ \gcc{}
above the  condensed surface.

The inset in Fig.~\ref{fig:structmag}
illustrates the structure of the envelopes of a neutron
star with $B=10^{14}$~G at the middle ages $t=10^{4\pm1}$~yr, around the
time of magnetic condensation. The two upper profiles are smooth,
without a condensed surface, and the two lower curves reveal a density
gap, which corresponds to an atmosphere and ocean above a dense solid
crust. As the star cools down, the ocean depth decreases. The
condensation is accompanied by increasing heat-transparency of the
envelope. For this reason, the decrease of the surface luminosity
stalls, and the two profiles in the middle (at $t=5$ kyr and 20 kyr,
before and after the condensation) are close to each other at low
densities. The slowdown of the luminosity decrease is clearly
distinguishable on the cooling curve for $B=10^{14}$~G at $t\sim10^4$ yr
in Figs.~\ref{fig:coolquant} and~\ref{fig:coolmagsf}.

\begin{figure}
\centering 
\includegraphics[width=\columnwidth]{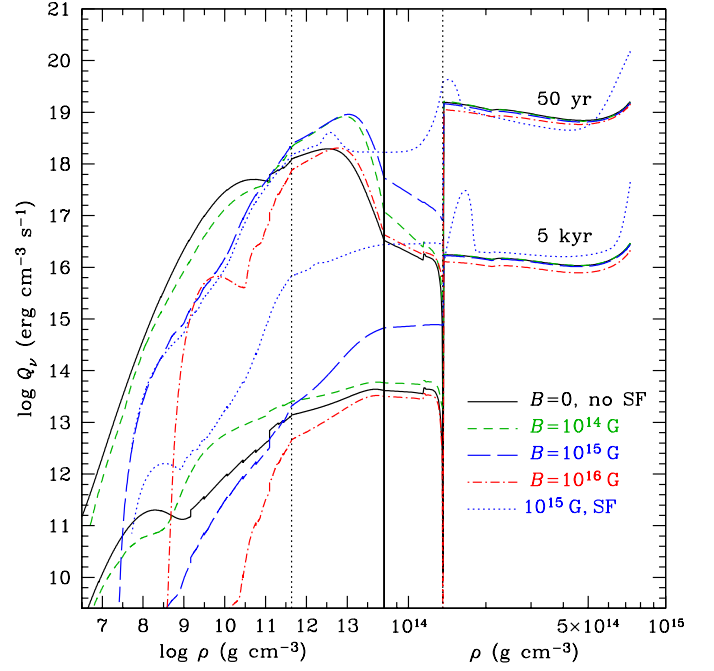}
\caption{Neutrino emission power density as a function of mass density at
ages 50 yr (upper curves) and 5 kyr (lower curves of each type) for
nonsuperfluid magnetars with $B=10^{14}$~G (short-dashed lines),
$10^{15}$~G (long-dashed lines), and $10^{16}$~G (dot-dashed lines) and
for a magnetar with $B=10^{15}$~G and nucleon superfluidity (dotted
lines), compared with neutrino emission of a nonmagnetic, nonsuperfluid
neutron star (solid lines).
The density scale is different in the left and right parts of the
Figure, separated by the vertical solid line. The vertical dotted lines
mark the outer and inner boundaries of the inner crust.
}
\label{fig:emismag} 
\end{figure}

\begin{figure}
\centering 
\includegraphics[width=\columnwidth]{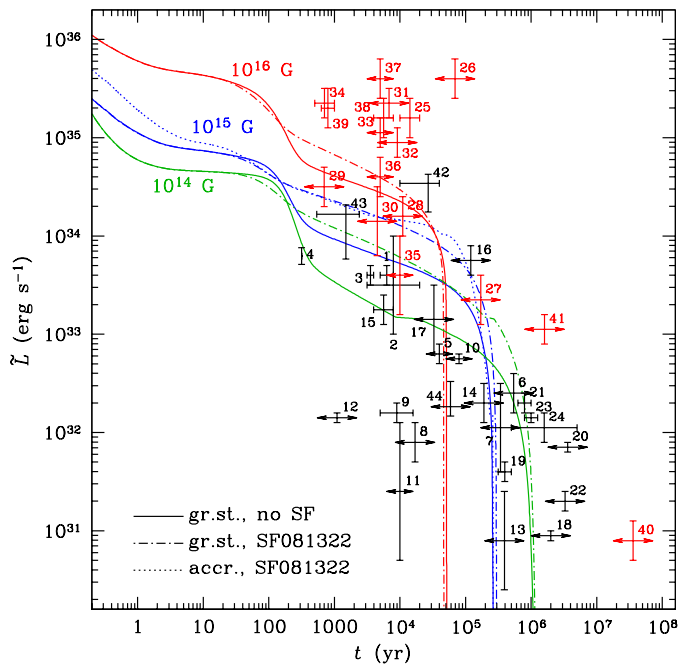}
\caption{Comparison of cooling of nonsuperfluid (solid lines) and
superfluid (dot-dashed lines) magnetars with superfluidity type SF081322
and magnetic fields $B=10^{14}$~G, $10^{15}$~G, and $10^{16}$~G. The
dotted line shows the joint effect of superfluidity and accreted envelope
for the magnetar with $B=10^{15}$~G.
}
\label{fig:coolmagsf} 
\end{figure}

Figure \ref{fig:emismag} shows neutrino emission rates inside
nonsuperfluid magnetars with $B=10^{14}$~G, $10^{15}$~G, and $10^{16}$~G
and a magnetar with superfluidity type SF081322 and $B=10^{16}$~G. For
reference, neutrino emission of the basic nonmagnetic neutron star model
is also shown. For ease of comparison with Fig.~\ref{fig:emission}, we have
chosen the same two ages, 50 yr and 5 kyr, and the same scale as in that
Figure. The most
remarkable difference is the enhanced emission in the inner crust for
$B=10^{14}$~G and $10^{15}$~G. The increase is due to the contribution
of the synchrotron neutrino emission. At the strongest considered field
$B=10^{16}$~G, however, the synchrotron contribution disappears, which
looks surprising at the first glance. Actually the synchrotron emission
is quenched at $T\ll T_B = T_\mathrm{cycl}/\sqrt{1+x_\mathrm{r}^2}$
 \citep{Bezchastnov_ea97}, 
where $T_\mathrm{cycl}$ is the electron cyclotron temperature defined in
Sect.~\ref{sect:physics} and $x_\mathrm{r}=\pF{,\ee}/\mel c$ is the
usual parameter of relativity. At
$B=10^{16}$~G the condition $T\ll T_B$ is fulfilled in the inner
crust. The quenching is described by a factor $S_\mathrm{BC}$ in
Sect.~2.4b of \citet{YKGH}, which is exponentially small at
$T_B/T\gg1$. 

We have seen that strong proton superfluidity in the core, accreted
envelopes, and superstrong magnetic fields enhance luminosities of
neutron stars at $t\sim10^2-10^4$ yr. One might expect that joint effect
of these factors would help to explain observations of the most luminous
magnetars. However, this is not the case. Figure \ref{fig:coolmagsf}
shows cooling of magnetars without and with strong proton superfluidity
along with an example of a superfluid magnetar with an accreted
envelope. We see that the stronger the magnetic field, the
smaller the additional enhancement of the photon luminosity due to the
superfluidity. Replacement of the iron envelope by an accreted envelope
does not produce any appreciable effect on the cooling of those
magnetars, whose luminosity has been already enhanced by a magnetic
field $B\gtrsim10^{15}$~G and strong proton superfluidity.

The cooling curves in Fig.~\ref{fig:coolmagsf} are compatible with the
observed luminosities and estimated ages of several magnetars without
involving heating mechanisms. However, most of the data on magnetars are barely
compatible with the theoretical cooling curves, and several objects (e.g., objects number 26,
4U 0142+614; 37, SGR 0526$-$66; 40, SGR 0526-66) are clearly incompatible with them. This indicates that
heating mechanisms are probably important for the thermal evolution of
magnetars, in agreement with previously published conclusions
\citep[e.g.,][]{Kaminker_ea09,Vigano_ea13,BeloborodovLi16}.
Nonetheless, Fig.~\ref{fig:coolmagsf} demonstrates that combined effects
of Landau quantization, magnetic condensation, and strong proton
superfluidity substantially reduce the discrepancies without
resorting to accreted envelopes of light elements, which would hardly
survive on the surface of magnetars, due to high temperatures
and bursting activity.

\section{Conclusions}
\label{sect:concl}

We have performed simulations of cooling of nonmagnetic neutron stars
and neutron stars with tangled superstrong magnetic fields. We have
studied the influence of various recent updates to microphysics input
and the effects brought about by superstrong magnetic fields of
magnetars. We treat the fully ionized
envelopes uniformly with the interior of the star while taking into
account the $T$-dependence of the EoS in the outer crust. We have
considered both neutron stars with ground-state composition and with
accreted envelopes.

We demonstrate that cooling simulations based on the approximation
of a quasi-stationary envelope, which extends to
$\rho_\mathrm{b}\sim10^{10}$ \gcc, are accurate on the timescales
$\gtrsim1$~yr, but not on the shorter timescales.
The accurate
treatment of the PBF neutrino emission, including the effect of
suppression of axial channel for triplet type of Cooper pairing of
neutrons, caused by contribution of the anomalous weak interaction
\citep{Leinson10}, can be important. In particular, it is shown that the
inclusion of this effect invalidates the tentative explanation of the
apparent rapid cooling of the Cas A NS by the PBF emission mechanism
\citep{Page_ea11,Ho_ea15}, in agreement
with \citet{Leinson16}. The enhancement of the Urca reactions by the
in-medium effects \citep{Voskresensky01,ShterninBaldo17} are equally
important. 

On the other hand, the latent heat of the crust, the in-medium effects
on baryon thermal conduction, the upgrade of BSk21 to BSk24 EoS model,
and the recent improvements in treatments of neutrino bremsstrahlung and
plasmon decay in the crust proved to have almost no effect on the
cooling. Allowance for the  electron-positron pairs, electron
degeneracy, Compton effect, and plasma cutoff in the treatment of
radiative opacities can noticeably (by $>10$\%) improve accuracy of the
simulations of thermal evolution only for very hot neutron stars, with
photon luminosities above $10^{36}$ erg s$^{-1}$, and are unimportant
for colder sources.

In agreement with previous studies, we find that accreted envelopes and
superstrong magnetic fields make the neutron-star envelope more
heat-transparent, which results in an increase of the surface luminosity
at the neutrino cooling stage and quick fading at a later photon cooling
stage. The cooling can be decelerated at the middle ages by a
combination of weak neutron and strong proton superfluidities. However,
the effects of the accreted envelopes and superstrong magnetic fields
are not additive. Thus the highest luminosity of a cooling magnetar
(without heating sources) at the middle ages $t\sim10^4$ yr is provided
by the superstrong magnetic field and superfluidity without an accreted
envelope. This maximum luminosity is still  not sufficient to explane
observations of the most luminous magnetars, which implies the importance
of heating. A study of the magnetar heating problem is beyond the scope
of the present paper.

We did not consider magnetic fields stronger than $10^{16}$~G for two
reasons; First, the strongest observed dipole component of a magnetic
field of a neutron star, evaluated from the star spindown rate, is
$2\times10^{15}$~G. While several times stronger small-scale fields look
plausible, orders of magnitude stronger fields do not. The second reason
is more fundamental. At $B>10^{16}$~G, a number of physical effects come
into play, which can be safely neglected at lower fields and have not
been included in the present study. Examples of
such effects are the $B$-dependence of the neutron drip transition
pressure \citep{Chamel_ea16}, a possible influence of the field on
nuclear shell energies and magic numbers, which may substantially change
the composition and structure of the crust
\citep{KondratievMC00,KondratievMC01,Stein_ea16}, and the shift of the
muon production threshold in the core \citep{SuhMathews01}.

\begin{acknowledgements}
We are grateful to Peter Shternin and Marcello Baldo for sharing
their results with us before publication. We thank the anonymous referee
for many valuable comments and suggestions, which helped us to improve
the paper. A.P.{} thanks Peter Shternin for critical remarks, which have
been taken into account in the final version of the text. The work of
A.P.{} was supported  by the Russian Science Foundation (grant
14-12-00316).
\end{acknowledgements}

\appendix

\section{Analytical parametrizations at $B=0$}
\label{sect:eosfits}

For modeling stellar structure and evolution, analytical
parametrizations of the quantities of astrophysical interest can be
useful. We have constructed such parametrizations for the BSk24 and APR
EoS models, following the approach previously developed by
\cite{HaenselPotekhin04} for Sly4 EoS and \citet{Potekhin_ea13} for
BSk21 EoS. The unified fits to pressure and energy density cover a broad
density range including the inner crust and the core. These unified fits
smear away all density discontinuities between layers of different
composition. Other quantities that are required in modeling neutron-star
structure and evolution, such as particle fractions, are given by
separate parametrizations for the inner crust and the core.

In the case of BSk24, the analytical fitting formulae for pressure $P$
as a function of mass density $\rho$, energy per baryon as a function of
baryon number density $\nbar$, number fractions of the electrons $Y_\ee$
in the core and free nucleons ($Y_\nn$, $Y_\pp$) in the inner crust,
sizes and shapes of nuclei in the inner crust, chemical potentials of
neutrons and protons in the inner crust and the core will be presented
in \citet{Pearson_ea17}. We do not reproduce them here, but refer the
reader to the 
web site\footnote{\url{http://www.ioffe.ru/astro/NSG/nseoslist.html}},
dedicated to Fortran implementations of the EoS fits mentioned in this
Appendix.

For the APR EoS, we have constructed
parametrizations of pressure and density in
exactly the same form as those presented by
\citet{HaenselPotekhin04} for the SLy4 EoS model of
\citet{DouchinHaensel01} in a wide density range
including the core and crust. Since APR
does not apply in the crust, the present unified fits rely on
the SLy4 EoS at $\nbar<\nbar_\mathrm{cc}$, where $\nbar_\mathrm{cc}$ is the
baryon number density at the crust-core transition
($\nbar_\mathrm{cc}=0.075959$ fm$^{-3}$ according to
\citealp{DouchinHaensel01}).

The parametrization of pressure $P$ as a function of mass density $\rho$
is
\begin{eqnarray}&&\!\!\!\!\!
  \log_{10}\left(\frac{P}{\mbox{dyn cm$^{-2}$}}\right) =
   \frac{6.22+6.121\xi+0.006035\,\xi^3}{(1+0.16354 \,\xi)
   \big[\ee^{4.73\, (\xi-11.5831)}+1\big]}
\nonumber\\&&\qquad
     + \frac{12.589+1.4365\,\xi}{\ee^{4.75\,(11.5756-\xi)}+1}
     + \frac{3.8175\,\xi-42.489}{\ee^{2.3\,(14.81 -\xi)}+1}
\nonumber\\&&\qquad
     + \frac{29.80-2.976\,\xi}{\ee^{1.99\, (14.93 -\xi)}+1}
     ,
\label{Pfit}
\end{eqnarray}
where $\xi\equiv \log_{10}(\rho/\textrm{g cm}^{-3})$. The
typical fit error of $P$ is $\approx1-2$\% for
$6\lesssim\xi\lesssim16$; the maximum error of $\lesssim
4$\% is due to the continuous fitting across the
discontinuity of density at the  phase boundary between the
crust and the core.

The mass density $\rho$ and the baryon number
density $\nbar$ are approximated as functions of each other by the
formulae
\begin{eqnarray}
   \frac{n_0}{\nbar} &=& 1+
     \left( 0.342 \nbar^{2.23}+2.2\times10^{-6} \nbar^{10.92}\right)
     \left(1+a_\nn^{-1}\right)^{-1}
\nonumber\\&&
     +  
    \frac{\nbar\,(1+a_\nn)^{-1}}{\big(8\times10^{-6} + 2.1\,\nbar^{0.585}\big)},
\quad
a_\nn=21.92\,\nbar^{0.3644}
     ~,
\label{rho-n}
\\
   \frac{n_0}{\nbar} &=& 1+
     \frac{0.173 n_0^{1.18}+9.97 n_0^{3.787}}{(1+2.634n_0)^3}\,
     \left(1+a_\rho^{-1}\right)^{-1}
\nonumber\\&&
     +  
    \frac{n_0\,(1+a_\rho)^{-1}}{8\times10^{-6} + 2.1\,n_0^{0.585}}\,
     ,
\quad
a_\rho=2.49\times10^{-5}\rho^{0.3982},
\label{n-rho}
\end{eqnarray}
where
$\nbar$ is in fm$^{-3}$, $\rho$ is in \gcc,
and $n_0=\rho/1.66\times10^{15}$.

The APR model predicts a phase transition at $\nbar\approx0.2$
fm$^{-3}$, which is accompanied by a drop of the charged 
fraction, as illustrated in Figs.~8 and 9 of \citet{AkmalPR98}. These results
can be approximately described by the formulae
\bea&&
   Y_\pp = 0.007+0.33\,\nbar
\quad 
   \mbox{at $\nbar_\mathrm{cc} < \nbar < 0.2$},
\\&&
   Y_\mu = \nbar/4-0.03
\quad 
   \mbox{at $0.12 < \nbar < 0.2$};
\\&&
   Y_\pp = \frac{0.62-\nbar}{1+2.5\,\nbar}
\quad\mbox{and~~}
   Y_\mu = 0.43\,Y_\pp
\quad 
   \mbox{at $\nbar > 0.2$},
\eea
where $\nbar$ is in fm$^{-3}$, 
$Y_\mu = 0$ at $\nbar < 0.12$,
and $Y_\ee=Y_\pp-Y_\mu$.

Calculations of thermal conductivity in the core, the neutron and proton
contributions to the specific heat, and some neutrino reaction rates 
require knowledge of the effective neutron and proton masses, $m_\nn^*$
and $m_\pp^*$. For the BSk family of models, the effective masses are
readily provided in an explicit analytical form as functions of the
baryon number density $\nbar$ and proton number fraction $Y_\pp$ by the
same generalized Skyrme parametrization that underlies the EoS
calculations, according to \citet{ChamelGP09}.  In the case of the APR
model, the effective masses for the AV18+UIX$^*$ forces that underlie
the EoS are given  at $\nbar\lesssim1$ fm$^{-3}$ by \citep{Baldo_ea14}
\bea
   m_\pp^* &=& 1-(1.56+1.31 Y_\pp-1.89Y_\pp^2)\,\nbar
\nonumber\\&&
   +
          (3.17+1.26Y_\pp-1.56Y_\pp^2)\,\nbar^2
\nonumber\\&&
   -
          (0.79+3.78Y_\pp-3.81Y_\pp^2)\,\nbar^3,
\\
   m_\nn^* &=& 1-(0.88+1.21Y_\pp+1.07Y_\pp^2)\,\nbar
\nonumber\\&&
   +     (1.64+2.06Y_\pp+2.87Y_\pp^2)\,\nbar^2
\nonumber\\&&
   - (0.78+0.98Y_\pp+1.62Y_\pp^2)\,\nbar^3.
\eea  

Calculations of electron thermal conductivity (\citealp{PotekhinPP15},
and references therein) and neutrino bremsstrahlung \citep{OfengeimKY14}
in the neutron star crust involve the proton equivalent size parameter
$x_\pp=R_\mathrm{nuc}/R_\mathrm{WS} =
\sqrt{5/3}\,R_\mathrm{ch}/R_\mathrm{WS}$, where
$R_\mathrm{nuc}$ is the nuclear radius in the model of the
rectangular profile of proton charge distribution,
$R_\mathrm{ch}$ is the root mean squared radius of the
proton charge distribution, and $R_\mathrm{WS}$ is the
equivalent Wigner-Seitz cell radius (the ion sphere radius).
For the BSk24 model, the size parameter is given as function
of $\nbar$ by the fit
\beq
   x_\pp =        \frac{0.1034+2.005\,\nbar^{0.575}
            }{
         1+72.7\,\nbar^{2.45}} + 1.06 Z \nbar^{2.61},
\eeq
where $\nbar$ is in fm$^{-3}$. This fit is accurate to within 1.8\%,
with typical residuals $\sim0.4$\%. For the NV model and the accreted
crust model, we use the estimate $R_\mathrm{nuc}\approx1.83 Z^{1/3}$~fm
by \citet{ItohKohyama83}. In the outer crust, we use the values of
$R_\mathrm{ch}$ provided by the \textsc{bruslib} database
\citep{Xu_ea13}.


\section{Equation of state of a strongly magnetized fermion plasma}
\label{sect:F_B}

The EoS of a fully degenerate nuclear matter in strong magnetic fields
was studied by \citet{Broderick00}, using the relativistic mean field
model. Here we apply an approximation,
which is quite accurate at $B\lesssim10^{16}$~G
and is not bound to using a specific field-theoretical model of
nucleon interactions. It relies on the fact that the magnetic field
$B\lesssim10^{16}$~G can be only weakly quantizing at
$\rho>\rho_\mathrm{nd}$, and for this reason its effects on the EoS in
the inner crust and the core can be treated perturbatively. We calculate
these effects approximately by adding to the free energy at $B=0$,
provided by a nonmagnetic EoS, the difference $\Delta F_B = F_B-F_0$
between the values for the given $B$ and for $B=0$, provided by a
generalization of the fully ionized electron-ion plasma model.

The free energy $F_B$ includes contributions due to the
electrons ($F_{B,\ee}$), positrons ($F_{B,\ee^+}$), neutrons
($F_{B,\mathrm{n}}$), nuclei in the
inner crust ($F_{B,Z,A}$), protons ($F_{B,\pp}$) in the
core (and in the deepest inner crust layers, where free protons are
present according to the BSk model), and muons ($F_{B,\mu}$) in the
core at the densities where free muons exist. For $F_{B,Z,A}$,
the known analytical formulae for the magnetized Coulomb crystal
\citep{PC13} are directly applicable. The other 
contributions are computed using the model of free Dirac fermions with
an addition of the anomalous magnetic moments. We
employ the thermodynamic relation
\beq
   F_{B,\alpha} = \mu_\alpha N_\alpha - P_\alpha V,
\label{F_B}
\eeq
where $V$ is the volume, $\mu_\alpha$ is an effective
chemical potential of the particles of type $\alpha$
($\alpha=\ee, ~\ee^+$, p, n, $\mu$), $n_\alpha = Y_\alpha
\nbar$ is their number density, $N_\alpha = n_\alpha V$, and
$P_\alpha$ is their partial pressure.

Anomalous magnetic moments affect the energies of relativistic particles
in a nontrivial way (see \citealp{Broderick00}).
Moreover, the $g$-factors of fermions are constant only in
the perturbative QED regime \citep[e.g.,][]{Schwinger88}, 
that is at $b_\alpha\ll1$, where
\beq
   b_\alpha = \frac{\hbar\omega_\alpha}{m_\alpha c^2}
      = \left(\frac{m_\ee}{m_\alpha}\right)^2
        \frac{B}{4.414\times10^{13}\mbox{~G}},
\eeq
$m_\alpha$ is the fermion mass, $\omega_\alpha=eB/m_\alpha c$,
and $e$ is the elementary charge. 
For the leptons, however, the anomalous magnetic
moments can be neglected (see \citealp{SuhMathews01}), while for
baryons we always have $b_\alpha\ll1$. 
Thus, in general, we have
\beq
  |\hat{g}_\alpha|\, b_\alpha \ll 1,
\label{small_gb}
\eeq
where
$\hat{g}_\alpha$ is the anomalous part of the g-factor
(for protons $\hat{g}_\pp=g_\pp-2=3.586$, for neutrons $\hat{g}_\nn=g_\nn=-3.826$,
and for leptons $\hat{g}_\alpha=g_\alpha-2\approx0$).
Under condition (\ref{small_gb}), the
energy due to the anomalous magnetic moment can be approximately treated
as an additive constant, positive of negative depending on the magnetic
moment direction. In
this approximation, the energy of a free charged relativistic fermion
in a magnetic field, including rest energy $m_\alpha c^2$, can be
written in a unified form,
\beq
   \epsilon_{\alpha,n,\sigma}(p_\|) = m_\alpha c^2 \sqrt{1
      + 2n b_\alpha 
      + (p_\|/m_\alpha c)^2} 
      - \sigma\,\hat{g}_\alpha\,\frac{\hbar\omega_\alpha}{4}
,\eeq
($\alpha=\ee$, $\ee^+$, $\mu$, p).
Here, $p_\|$ is the momentum along the field and
$n=n_\rho+(1+\sigma)/2$ is the conventional Landau quantum number,
$n_\rho=0,1,2,\ldots$ being the radial Landau quantum number;
$\sigma=\pm1$ controls the spin projection on the magnetic field.  A
straightforward generalization of Eqs.~(51) and (52) of \citet{PC13}
then reads
\bea
   n_\alpha &=& 
  \frac{1}{\pi^{3/2}\,\am^2\,\lambda_\alpha}
   \sum_{\sigma=\pm1} \,\sum_{n=(1+\sigma)/2}^{\infty}
    (1+2nb_\alpha )^{1/4}\,
     \frac{\partial I_{1/2}(\chi_{\alpha,n,\sigma},\tau_{\alpha,n})}{
     \partial \chi_{\alpha,n,\sigma}},
\nonumber\\[-2ex]&&
\label{densmag}
\\
    P_\alpha &=&
  \frac{\kB T}{\pi^{3/2}\,\am^2\,\lambda_\alpha}
   \sum_{\sigma=\pm1} \,\sum_{n=(1+\sigma)/2}^{\infty}
    (1+2nb_\alpha )^{1/4}\, I_{1/2}(\chi_{\alpha,n,\sigma},\tau_{\alpha,n}),
\nonumber\\[-2ex]&&
\label{presmag}
\eea
where 
$\lambda_\alpha = \sqrt{{2\pi \hbar^2}/{m_\alpha\kB T}}$
is the thermal de Broglie length, $\am=\sqrt{\hbar c/eB}$ is the
magnetic length,
\bea&&
\tau_{\alpha,n} = \frac{\kB T}{m_\alpha c^2 \sqrt{1+2nb_\alpha }},
\\&&
   \chi_{\alpha,n,\sigma} = \frac{\mu_\alpha -
   m_\alpha c^2 \sqrt{1+2nb_\alpha }}{\kB T}
   + \frac{\sigma\hat{g}_\alpha\zeta_\alpha}{4},
\quad
   \zeta_\alpha\equiv\frac{\hbar\omega_\alpha}{\kB T},
\eea
and
\beq
   I_\nu(\chi,\tau) \equiv \int_0^\infty
  \frac{ x^\nu\,(1+\tau x/2)^{1/2}
    }{ \exp(x-\chi)+1 }\,{\dd}x
\label{I_nu}
,\eeq
which is the relativistic Fermi-Dirac integral.

For neutrons, in the same approximation, the energy is
\beq
   \epsilon_{\nn,\sigma} = \sqrt{(m_\nn c^2)^2
      + (pc)^2} 
      - \sigma \hat{g}_\nn\,\frac{\hbar\omega_\nn}{4},
\eeq
where $p$ is the momentum.
This leads to
\bea
   n_\alpha &=& \frac{2}{\sqrt{\pi}\,\lambda_\alpha^3}
    \sum_{\sigma=\pm1}
   \left[ I_{1/2}(\chi_{\alpha,0,\sigma},\tau_{\alpha,0})
   + \tau_{\alpha,0} I_{3/2}(\chi_{\alpha,0,\sigma},\tau_{\alpha,0}) \right],
\nonumber\\[-2ex]&&
\label{n_e} 
 \\
 P_\alpha &=&
 \frac{4}{3\sqrt\pi}\,\frac{\kB T }{ \lambda_\alpha^3}
    \sum_{\sigma=\pm1}
   \Big[ I_{3/2}(\chi_{\alpha,0,\sigma},\tau_{\alpha,0})
   + \frac{\tau_{\alpha,0}}{ 2}
     I_{5/2}(\chi_{\alpha,0,\sigma},\tau_{\alpha,0})
    \Big].
\nonumber\\[-1ex]&&
\label{P_e}
\eea
Equations (\ref{n_e}) and (\ref{P_e}) are valid not only for neutrons
($\alpha=\nn$), but also for
other fermions in a nonquantizing magnetic field or at $B=0$ (in the
latter case, $\sum_\sigma$ amounts to the factor 2),
so they yield $F_0$.

At given $n_\alpha$, $T$, and $B$, we find $\mu_\alpha$ by
numerical inversion of Eqs.~(\ref{densmag}) or (\ref{n_e}). 
Then the above
expressions for $n_\alpha$ and $P_\alpha$ provide
the partial free energy
$F_{B,\alpha}$, from which we can derive
the magnetic 
contributions of the fermions $\alpha$ into the total free energy
$F$, entropy $S= -(\partial F / \partial T )_V$, internal energy
$U=F+TS$, heat capacity at constant volume $C_V=(\partial S/\partial\ln
T)_V$, derivatives of pressure 
$
   (\partial P/\partial T)_V
$ 
and
$
   (\partial P/\partial V)_T
$, 
heat capacity at constant pressure $C_P=C_V-(\partial
P/\partial T)_V^2/(\partial P/\partial V)_T$, and so on.

Analytical approximations for the Fermi-Dirac integrals
$I_\nu(\chi,\tau)$ (see \citealp{PC13} and references
therein) provide
$n_\alpha(\mu_\alpha,T)$, $P_\alpha(\mu_\alpha,T)$, and
consequently
$F_{B,\alpha}(\mu_\alpha,T)$ in an explicit form.
Using relations
\bea&&
   \left(\frac{\partial f(\mu_\alpha,T)}{\partial T}\right)_V =
   \left(\frac{\partial f}{\partial T}\right)_{\!\mu_\alpha} \!\!
   + \left(\frac{\partial f}{\partial \mu_\alpha}\right)_T
   \, \left(\frac{\partial \mu_\alpha}{\partial T}\right)_V,
\label{dfdT}
\\&&
   \left(\frac{\partial f(\mu_\alpha,T)}{\partial V}\right)_T \!\! =
   \left(\frac{\partial f}{\partial T}\right)_{\mu_\alpha}
   \, \left(\frac{\partial \mu_\alpha}{\partial V}\right)_T
\\&&
   \left(\frac{\partial \mu_\alpha}{\partial T}\right)_V =
   - \frac{(\partial n_\alpha/\partial T)_{\mu_\alpha}
   }{
   (\partial n_\alpha/\partial\mu_\alpha)_T},
\quad
   \left(\frac{\partial \mu_\alpha}{\partial V}\right)_T =
   - \frac{n_\alpha}{V}
   \left(\frac{\partial n_\alpha}{\partial
   \mu_\alpha}\right)_T^{-1},
\label{deriv_mu}
\eea
we can now write explicit analytical
approximations to first, second, and mixed partial
derivatives of $F_{B,\alpha}$ over $V$ and $T$,
which provide magnetic contributions to 
the above-mentioned thermodynamic functions.

In the particular case of nonrelativistic nondegenerate particles with
spin 1/2, we have
\beq
   F_{B,\alpha} = N_\alpha m_\alpha c^2 +
   F_{B,\alpha}^\mathrm{kin} + F_{B,\alpha}^\mathrm{spin},
\eeq
where
\beq
  F_{B,\alpha}^\mathrm{kin} = N_\alpha \kB T \left[
    \ln(2\pi\lambda_\alpha\am^2) +
    \ln\left(\ee^{\zeta_\alpha/2}-\ee^{-\zeta_\alpha/2}\right) - 1
    \right],
\eeq
which is the kinetic contribution for charged particles,
\beq
  F_{B,\alpha}^\mathrm{kin} = N_\alpha \kB T \left[
    \ln(2\pi\lambda_\alpha^3) - 1
    \right]
\eeq
which is the kinetic contribution for neutrons (or in nonquantizing magnetic
field), and
\beq
   F_{B,\alpha}^\mathrm{spin} = - \,N_\alpha \kB T \left[
    \frac{g_\alpha\zeta_\alpha}{4}
    + \ln\left(\frac{1-\ee^{-g_\alpha\zeta_\alpha}}{
      1-\ee^{-g_\alpha\zeta_\alpha/2}} \right)
      \right]
\eeq
which is the magnetic moment contribution in both cases.
The pressure of nondegenerate 
particles is not affected by the magnetic field $(P_\alpha = n_\alpha\kB
T)$, but the internal energy
$U_\alpha=N_\alpha m_\alpha c^2 +
U_\alpha^\mathrm{kin}+U_\alpha^\mathrm{spin}$ and heat capacity 
$C_{V\alpha}=C_{V\alpha}^\mathrm{kin}+C_{V\alpha}^\mathrm{spin}$
are affected by the $B$-dependent spin contributions
\beq
   \frac{U_\alpha^\mathrm{spin}}{N_\alpha \kB T} =
   - \frac{g_\alpha\zeta_\alpha}{4}\,
   \mathrm{tanh}\frac{g_\alpha\zeta_\alpha}{4},
\quad
  \frac{C_{V\alpha}^\mathrm{spin}}{N_\alpha \kB} =
   \left(\frac{g_\alpha\zeta_\alpha/4}{
     \mathrm{cosh}(g_\alpha\zeta_\alpha/4)} \right)^2
    \!\!,
\eeq
and for charged particles also by the $B$-dependent kinetic
contributions,
\beq
   \frac{U_\alpha^\mathrm{kin}}{N_\alpha \kB T} =
   \frac12 + \frac{\zeta_\alpha}{\ee^{\zeta_\alpha}-1}
   + \frac{\zeta_\alpha}{2},
\quad
  \frac{C_{V\alpha}^\mathrm{kin}}{N_\alpha \kB} =
    \frac12 + \left(\frac{\zeta_\alpha}{\ee^{\zeta_\alpha}-1} \right)^2
    \!\!.
\eeq

In the opposite case of strongly degenerate fermions
($\mu_\alpha-m_\alpha c^2\gg\kB T$), one can use the
Sommerfeld expansion (e.g., \citealp{Chandra}, Chapter~X),
\bea
     I_\nu(\chi,\tau) &=&
        \frac{2^{-1/2}}{\tau^{\nu+1}}\left(
        \mathcal{I}_\nu^{(0)}(\tilde\mu)
          +\frac{\pi^2\tau^2}{6} \mathcal{I}_\nu^{(2)}(\tilde\mu)
    +   \frac{7\pi^4\tau^4}{360} \mathcal{I}_\nu^{(4)}(\tilde\mu)
    + \ldots
          \right),
\nonumber\\&&
\label{Sommer}
\eea
where $\tilde\mu\equiv\chi\tau$,
\bea&&
\mathcal{I}_{1/2}^{(0)}(\tilde\mu)
 = \frac{\tilde{x}\,(1+\tilde{\mu})-\ln(\tilde{x}+1+\tilde{\mu}) }{2}
,
\label{I12}
\\&&
\mathcal{I}_{3/2}^{(0)}(\tilde\mu)
 = \frac{\tilde{x}^3}{3} - \mathcal{I}_{1/2}^{(0)}(\tilde\mu)
,
\\&&
\mathcal{I}_{5/2}^{(0)}(\tilde\mu)
 = \tilde{x}^3\,\frac{1+\tilde{\mu}}{4} - \frac{2 \tilde{x}^3}{3}
  + \frac54\,\mathcal{I}_{1/2}^{(0)}(\tilde\mu)
,
\label{I52}
\\&&
   \mathcal{I}_\nu^{(k+1)}(\tilde\mu)
   =
    \frac{\dd \mathcal{I}_\nu^{(k)}(\tilde\mu)}{\dd\tilde\mu} ,
\label{dIdmu}
\eea
and
$\tilde{x} \equiv \sqrt{\tilde\mu(2+\tilde\mu)}$. At a fixed
$n_\alpha$, to the lowest order in $T^2$, 
Eqs.~(\ref{F_B}), (\ref{densmag}), and (\ref{Sommer}) yield
\bea
   \Delta\tilde\mu_\alpha \equiv
   \frac{\mu_\alpha - \EF{\alpha}}{m_\alpha c^2} &=& 
   - \frac{\pi^2\tau_{\alpha,0}^2}{6}\,
   \frac{\sum_{\sigma,n}(1+2nb_\alpha )^{-1/2}
   \mathcal{I}_{1/2}^{(3)}(\tilde\mu_{\alpha,n,\sigma})}{
   \sum_{\sigma,n}(1+2nb_\alpha )^{1/2}
   \mathcal{I}_{1/2}^{(2)}(\tilde\mu_{\alpha,n,\sigma})},
\nonumber\\&&
\eea
where summation indices run over the same values as before
 ($\sigma=\pm1$; $n=0,1,2,\ldots$ for $\sigma=-1$ and $n=1,2,3,\ldots$
 for $\sigma=1$),
\beq
   \tilde\mu_{\alpha,n,\sigma} \equiv 
   \frac{\EF{\alpha}}{m_\alpha c^2} - \sqrt{1+2nb_\alpha }
    + \frac{\sigma\hat{g}_\alpha b_\alpha}{4},
\eeq
and
$\EF{\alpha}$ is the Fermi energy, which is found from the
condition
\beq
   n_\alpha = \frac{m_\alpha c}{2\pi^2\hbar\am^2}\sum_{\sigma,n}
   \sqrt{1+2nb_\alpha}\,
   \mathcal{I}_{1/2}^{(1)}(\tilde\mu_{\alpha,n,\sigma}).
\eeq
The pressure is given, to the same order of approximation, by
\beq
   P_\alpha \approx P_\alpha^{(0)} + P_\alpha^{(1)}
,\eeq
where $P_\alpha^{(0)}=P(T=0)$ and $P_\alpha^{(1)}\propto T^2$ are given
by
\bea
   P_\alpha^{(0)} &=&
   \frac{m_\alpha^4 c^5}{\hbar^3}\,
   \frac{b_\alpha}{2\pi^2}\,\sum_{\sigma, n}
   (1+2nb_\alpha)\mathcal{I}_{1/2}^{(0)}(\tilde\mu_{\alpha,n,\sigma}),
\\
   P_\alpha^{(1)} &=& \frac{m_\alpha^4 c^5}{\hbar^3}\,
   \frac{b_\alpha}{2\pi^2}\,\sum_{\sigma, n}
   \Big[ \Delta\tilde\mu_\alpha 
   (1+2nb_\alpha)\,\mathcal{I}_{1/2}^{(1)}(\tilde\mu_{\alpha,n,\sigma})
\nonumber\\&&
   + \frac{\pi^2\tau_{\alpha,0}^2}{6}\,
   \mathcal{I}_{1/2}^{(2)}(\tilde\mu_{\alpha,n,\sigma})
   \Big].
\eea

For the degenerate neutrons, in the same approximation, 
\req{Sommer} and 
Eqs.~(\ref{F_B}), (\ref{n_e}), and (\ref{P_e}) yield
\beq
   \Delta\tilde\mu_\alpha = 
   - \frac{\pi^2\tau_{\alpha,0}^2}{6}\,
   \frac{\sum_\sigma 
   \left[\mathcal{I}_{1/2}^{(2)}(\tilde\mu_{\alpha,0,\sigma})
   + \mathcal{I}_{3/2}^{(2)}(\tilde\mu_{\alpha,0,\sigma}) \right]}{
   \sum_{\sigma} \left[
   (\mathcal{I}_{1/2}^{(1)}(\tilde\mu_{\alpha,0,\sigma})
   + \mathcal{I}_{3/2}^{(1)}(\tilde\mu_{\alpha,0,\sigma})
   \right]},
\eeq
where $\EF{\alpha}$ is found from the
condition
\beq
   n_\alpha = \frac{1}{2\pi^2}\left(\frac{\hbar}{m_\alpha c}\right)^3
     \sum_\sigma \left[
     \mathcal{I}_{1/2}^{(0)}(\tilde\mu_{\alpha,0,\sigma})
   + \mathcal{I}_{3/2}^{(0)}(\tilde\mu_{\alpha,0,\sigma}) \right].
\eeq
Then the two lowest-order contributions to pressure are
\bea
   P_\alpha^{(0)} 
   &=& \frac{m_\alpha^4 c^5}{3\pi^2\hbar^3}\,
   \sum_{\sigma}
   \left[ \mathcal{I}_{3/2}^{(0)}(\tilde\mu_{\alpha,0,\sigma})
   + \frac12\,\mathcal{I}_{5/2}^{(0)}(\tilde\mu_{\alpha,0,\sigma})
   \right],
\\
   P_\alpha^{(1)} &=& \frac{m_\alpha^4 c^5}{3\pi^2\hbar^3}\,
   \Delta\tilde\mu_\alpha \sum_{\sigma}
   \left[ \mathcal{I}_{3/2}^{(1)}(\tilde\mu_{\alpha,0,\sigma})
   + \frac12\,\mathcal{I}_{5/2}^{(1)}(\tilde\mu_{\alpha,0,\sigma})
   \right]
\nonumber\\&&
   +   
    \frac{m_\alpha^2 c}{18\hbar^3}\,(\kB T)^2
   \sum_{\sigma}
   \left[ \mathcal{I}_{3/2}^{(2)}(\tilde\mu_{\alpha,0,\sigma})
   + \frac12\,\mathcal{I}_{5/2}^{(2)}(\tilde\mu_{\alpha,0,\sigma})
   \right].
\eea

These approximations give the free energy
with the lowest-order $T$-dependent terms in the form
\bea&&
   F_{B,\alpha} = F_{B,\alpha}^{(0)} + F_{B,\alpha}^{(1)},
\\&&
   F_{B,\alpha}^{(0)} =
   N_\alpha \EF{\alpha} 
    - V P_\alpha^{(0)}  = F_{B,\alpha}\Big|_{T=0},
\\&&
   F_{B,\alpha}^{(1)} = N_\alpha m_\alpha c^2\Delta\tilde\mu_\alpha
    - P_\alpha^{(1)} V \propto T^2,
\eea
from which the first- and second-order thermodynamic functions are
easily derived. They take particularly simple forms for the
nonrelativistic fermions, for which $\EF{\alpha}-m_\alpha c^2 \ll
m_\alpha c^2$.  Then we can consider $\tilde\mu\ll1$ in
Eqs.~(\ref{I12})\,--\,(\ref{dIdmu}), and they simplify to
\beq
   \mathcal{I}_{1/2}^{(0)} \approx \frac{\tilde{x}^3}{3},
\quad\!
   \mathcal{I}_{3/2}^{(0)}  \approx \frac{\tilde{x}^5}{10},
\quad\!
   \mathcal{I}_{5/2}^{(0)}  \approx  \frac{\tilde{x}^7}{28},
\quad\!
\mathcal{I}_\nu^{(k+1)}\approx \frac{1}{\tilde{x}}\,
        \frac{\dd \mathcal{I}_\nu^{(k)}}{
    \dd\tilde{x}}.
\eeq
Taking into account that for the baryons $b_\alpha\ll1$, we
find, for example, that the Fermi energy of strongly degenerate neutrons
$\EF{\nn}(n_\nn,B)\approx\EF{\nn}(n_\nn,0)\left[ 1-(g_\nn
\hbar\omega_\nn/8\EF{\nn})^2 \right]$, and that 
the temperature corrections
$\Delta\tilde\mu_\nn$ and $P_\nn^{(1)}$ are shifted relative to the
nonmagnetic expressions [Eq.~(6) in \citealp{PC10}] by the same
fractional order of magnitude $\sim(\hbar\omega_\nn/\EF{\nn})^2$.
Analogous corrections for
nonrelativistic protons are, by order of magnitude,
$\sim(\hbar\omega_\pp/\EF{\pp})^2$. Thus the contributions to the
thermodynamic functions (in particular, heat capacity) of strongly
degenerate baryons, caused by their anomalous magnetic moments, prove to
be unimportant, in contrast to the case of nondegenerate baryons.

\newcommand{\artref}[4]{{#4}, {#1}, {#2}, #3}
\newcommand{\AandA}[3]{\artref{A\&A}{#1}{#2}{#3}}
\newcommand{\AN}[3]{\artref{Astron.\ Nachr.}{#1}{#2}{#3}}
\newcommand{\ApJ}[3]{\artref{ApJ}{#1}{#2}{#3}}
\newcommand{\ApJS}[3]{\artref{ApJS}{#1}{#2}{#3}}
\newcommand{\ApSS}[3]{\artref{Ap\&SS}{#1}{#2}{#3}}
\newcommand{\ARAA}[3]{\artref{ARA\&A}{#1}{#2}{#3}}
\newcommand{\MNRAS}[3]{\artref{MNRAS}{#1}{#2}{#3}}
\newcommand{\NP}[4]{\artref{Nucl.\ Phys. #1}{#2}{#3}{#4}}
\newcommand{\PL}[4]{\artref{Phys.\ Lett. #1}{#2}{#3}{#4}}
\newcommand{\PR}[4]{\artref{Phys.\ Rev. #1}{#2}{#3}{#4}}
\newcommand{\PRL}[3]{\artref{Phys.\ Rev.\ Lett.}{#1}{#2}{#3}}
\newcommand{\RMP}[3]{\artref{Rev.\ Mod.\ Phys.}{#1}{#2}{#3}}
\newcommand{\SSRv}[3]{\artref{Space Sci.\ Rev.}{#1}{#2}{#3}}

\end{document}